\documentclass{aa}
\usepackage{natbib}
\usepackage{graphics}
\usepackage{amssymb}
\begin{document}
\title{Imaging chemical differentiation around the low-mass protostar
L483-mm} 
\author{J.K. J{\o}rgensen} 
\institute{Leiden Observatory, P.O. Box 9513, NL-2300 RA Leiden, Netherlands} 
\offprints{Jes K.\,J{\o}rgensen} 
\mail{joergensen@strw.leidenuniv.nl} 
\date{Received 11 February 2004 / Accepted 14 May 2004}

\abstract{This paper presents a millimeter wavelength
aperture-synthesis study of the spatial variations of the chemistry in
the envelope around the deeply embedded low-mass protostar L483-mm on
$\sim$1000~AU (5\arcsec) scales. Lines of 8 molecular species
including CN, C$^{18}$O, CS, C$^{34}$S, HCN, H$^{13}$CN, HCO$^+$ and N$_2$H$^+$
have been observed using the Owens Valley Radio Observatory Millimeter
Array. Continuum emission at 2.7-3.4~millimeter is well-fit by an
envelope model based on previously reported submillimeter continuum
images down to the sensitivity of the interferometer without
introducing a disk/compact source, in contrast to what is seen for
other protostellar objects. A velocity gradient in dense material
close to the central protostar is traced by HCN, CS and N$_2$H$^+$, and is
perpendicular to the large-scale CO outflow, with a pattern consistent
with rotation around a $\sim$1~$M_\odot$ central object. Velocity
gradients in the propagation direction of the outflow suggest a clear
interaction between the outflowing material and ``quiescent''
core. Significant differences are observed between the emission
morphologies of various molecular species. The C$^{18}$O
interferometer observations are fit with a ``drop'' abundance profile
where CO is frozen-out in a region of the envelope with temperatures
lower than 40~K and densities higher than 1.5$\times 10^5$~cm$^{-3}$,
which is also required to reproduce previously reported single-dish
observations. The N$_2$H$^+$ emission strongly resembles that of NH$_3$
and is found to be absent toward the central continuum source. This is
a direct consequence of the high CO abundances in the inner region as
illustrated by a chemical model for the L483 envelope. The observed CN
emission forms a spatial borderline between the outflowing and
quiescent material probed by, respectively, HCO$^+$ and N$_2$H$^+$, and
also shows intermediate velocities compared to these two species. A
scenario is suggested in which CN is enhanced in the walls of an
outflow cavity due to the impact of UV irradiation either from the
central protostellar system or related to shocks caused by the
outflow.

} \maketitle

\section{Introduction}
The chemistry of star-forming regions shows a richness and complexity
reflecting large variations in the physical conditions found in these
environments. For example, the thermal evolution of the pre- and
protostellar cores results in evaporation and freeze-out of molecules
illustrating the important interplay between the solid-state and
gas-phase chemistry. Recent single-dish surveys of low-mass protostars
\citep{jorgensen02,paperii,schoeier02,maret04} have illustrated that
the chemistry may be severely affected by the thermal history of
cores, with radial variations of the temperature due to the heating
from the central protostar. This paper presents high-resolution
millimeter wavelength aperture-synthesis observations of a wide range
of different molecules toward the embedded low-mass protostar
\object{L483-mm}. In combination with a detailed radiative transfer
model \citep{jorgensen02,paperii} these high-resolution observations
make it possible to resolve and address the spatial variation of the
chemistry in the protostellar envelope on $\sim$1000~AU (5\arcsec)
scales. Only a few protostars have previously been studied in this
chemical detail and the combination with line radiative transfer
models and information about the larger scale structure from
single-dish observations makes a unique discussion about the chemical
differentiation at varying temperatures and densities possible.

Recent single-dish continuum studies of embedded protostars have
proved useful to establish the physical structure of the protostellar
envelopes. A detailed description of the density and temperature
variation is crucial for calculations of the molecular excitation and
for the interpretation of the chemical small-scale structure from high
resolution interferometer observations. Typically the envelope density
and temperature structure can be constrained from radiative transfer
modeling of submillimeter continuum observations
\citep[e.g.,][]{jorgensen02,shirley02} or from infrared extinction
measurements \citep[e.g.,][]{alves01,harvey01}. The detailed profiles
can subsequently be used as input for Monte Carlo radiative transfer
modeling of single-dish line emission to constrain the molecular
abundances
\citep[e.g.,][]{bergin02,jorgensen02,paperii,schoeier02}. The models
can also be used to predict images of both continuum and line emission
that can be directly compared to interferometer observations
constraining the envelope structure down to scales of a few hundred AU
\citep[e.g.,][]{n1333i2art,hotcorepaper}. The caveat of the
single-dish studies is their low spatial resolution of 10-15\arcsec\
or worse, which is why these observations predominantly are sensitive
to material on larger scales ($\gtrsim$2000~AU) unless high-excitation
lines are observed. Recent interferometric studies of millimeter
continuum emission around low-mass protostars
\citep[e.g.,][]{hogerheijde99,looney00,harvey03,n1333i2art,hotcorepaper}
have illustrated the potential for probing the small-scale physical
structure of the envelopes and constrain the presence of unresolved
emission, possibly originating in circumstellar disks in these deeply
embedded stages.

Images of molecular line emission at similar resolutions can be used
to discuss the detailed envelope chemical structure. For example,
\cite{n1333i2art} reported high-resolution observations of a range of
molecular species toward the class 0 protostar,
\object{NGC~1333-IRAS2A}. The 3-6\arcsec\ (600-1200~AU) observations
were interpreted in the context of envelope models constrained by
single-dish continuum maps and multi-transition line observations from
the survey by \cite{jorgensen02,paperii}. It was found that the
single-dish envelope model could be successfully extrapolated to the
smaller scales, lending further credibility to the approach and
derived physical and chemical properties.

\object{L483-mm} (\object{IRAS~18148-0440}; in the following just
L483) is similar to \object{NGC~1333-IRAS2A} being a deeply embedded,
low-mass protostar with a low bolometric temperature of $\approx
50$~K. In contrast to \object{NGC~1333-IRAS2A}, it shows a remarkable,
asymmetric structure in the SCUBA maps. This asymmetry is likely to be
the cause of the rather flat density distributions ($n\propto r^{-p}$)
found in radiative transfer modeling of the continuum emission by
\cite{jorgensen02} ($p=0.9$) and \cite{shirley02} ($p=1.2$). The
envelope parameters for L483 are given in Table~\ref{l483_params}.
\begin{table}
\caption{Parameters for L483 from \cite{jorgensen02}.}\label{l483_params}
\begin{tabular}{ll}\hline\hline
Distance, $d$                        & 200~pc               \\
$L_{\rm bol}$                        &  9~$L_\odot$         \\
$T_{\rm bol}$                        & 50~K                 \\ \hline
\multicolumn{2}{l}{\sl Envelope parameters:}                \\ \hline
Inner radius ($T=250$~K), $R_i$      & 9.9 AU               \\
Outer radius, $R_{\rm 10 K}$         & 1.0$\times 10^4$ AU        \\
Density at 1000~AU, $n$(H$_{2}$)        & 1.0$\times 10^6$ cm$^{-3}$ \\
Slope of density distribution, $p$   & 0.9                  \\
Mass, $M_{\rm 10 K}$                 & 4.4~$M_\odot$        \\\hline
\end{tabular}
\end{table}

\cite{park00} reported BIMA observations of HCO$^+$ 1--0 and
C$_3$H$_2$ $2_{12}-1_{01}$ toward L483. They found two characteristic
velocity gradients in the two species: an east-west velocity gradient
in HCO$^+$ coincident with the larger scale CO outflow and a
north-south velocity gradient in C$_3$H$_2$ which they associated with
global contraction of an envelope of a few thousand AU size. L483 has
also been mapped in NH$_3$, most recently by \cite{fuller00} who
presented NH$_3$ maps from the VLA. They found that the overall
asymmetry observed with SCUBA also shows up in the NH$_3$ maps, but
with NH$_3$ lacking emission close to the protostar, which they
ascribed to optical depth effects. They also noted a characteristic
velocity pattern across the maps and suggested the presence of infall
close to the central protostar.

In addition to clear signatures in CO emission \citep{fuller95,
hatchell99, tafalla00}, an outflow driven by L483 is also seen through
near-infrared emission, possibly caused by scattered emission in the
outflow cavities \citep{hodapp94,fuller95}. The absence of significant
enhancements of CH$_3$OH and SiO, as found in other outflows, led
\cite{tafalla00} to suggest that the L483 outflow is more evolved than
those from other class 0 objects. Since the driving source itself
appears deeply embedded \citep{fuller95,fuller00},
\citeauthor{tafalla00} therefore suggested that L483 is in transition
from the class 0 to the class I stage.

This paper presents high resolution 3~mm interferometer observations
of a wide range of molecular species toward L483. The main objective
is to address whether the observed trends in the single-dish survey by
\cite{paperii} can be related to the variations in the chemistry in
the L483 envelope. The paper is laid out as follows:
Sect.~\ref{observations} describes the details of the
observations. Sect.~\ref{continuum} discusses the continuum maps of
L483 and compares to the predictions from the envelope model based on
the SCUBA observations from \cite{jorgensen02}. Sect.~\ref{lineobs}
presents the line observations. Sect.~\ref{discussion} brings together
the results for the line emission and discusses the implications for
the chemistry in the environment of L483.

\section{Observations}\label{observations}
\object{L483-mm} ($\alpha_{2000}=18^{\rm h}17^{\rm m}29\fs8$,
$\delta_{2000}=-04^\circ39'38\farcs3$) was observed using the Owens
Valley Radio Observatory (OVRO) Millimeter Array\footnote{The Owens
Valley Millimeter Array is operated by the California Institute of
Technology under funding from the US National Science Foundation
(grant no. AST-9981546).} from October 2001 to February 2003 in three
settings including HCN, H$^{13}$CN and HCO$^+$ (3.4~mm), N$_2$H$^+$,
CS and C$^{34}$S (3.2~mm), and C$^{18}$O and CN (2.7~mm) as summarized
in Table~\ref{datasum}. Each of the settings was observed in the C and
E configurations providing projected baselines ranging from 3 to
45~k$\lambda$. Compared to the BIMA HCO$^+$ maps presented by
\cite{park00}, the $(u,v)$ coverage of the tracks in this paper
includes longer baselines but not the shortest spacings. These
observations are therefore less sensitive to extended emission but
provide higher resolution. The correlator gives spectral resolutions
of $\approx$~0.15-0.2~km~s$^{-1}$ over 128 channels covering each line
with RMS noise levels of 0.1~Jy~beam$^{-1}$ per channel. The complex
gains were calibrated by regular observations of the nearby quasar,
j1743-038, approximately every 20-25 minutes. The bandpass was
calibrated by observations of strong quasars, the fluxes using the
same quasars as secondary calibrators and Uranus and Neptune as
primary calibrators. Calibration of the data was performed using the
MMA reduction package \citep{scoville93}.
\begin{table}[!htb]\centering
\caption{Summary of the observed lines.}\label{datasum}
\begin{tabular}{lll}\hline\hline
Molecule & Transition    & Frequency  \\ \hline
H$^{13}$CN   & 1--0$^a$      & \phantom{1}86.3402 \\
HCN      & 1--0$^a$      & \phantom{1}88.6318 \\
HCO$^+$   & 1--0          & \phantom{1}89.1885 \\[0.5ex] 
N$_2$H$^+$   & 1--0$^a$      & \phantom{1}93.1737 \\
C$^{34}$S    & 2--1          & \phantom{1}96.4129 \\
CS       & 2--1          & \phantom{1}97.9810 \\[0.5ex]
C$^{18}$O    & 1--0          & 109.7822           \\
CN       & 1--0$^a$      & 113.4910           \\\hline
\end{tabular}

Notes: $^a$Hyperfine splitting observed.
\end{table}

\section{Continuum emission}\label{continuum}
\begin{figure}
\resizebox{\hsize}{!}{\includegraphics{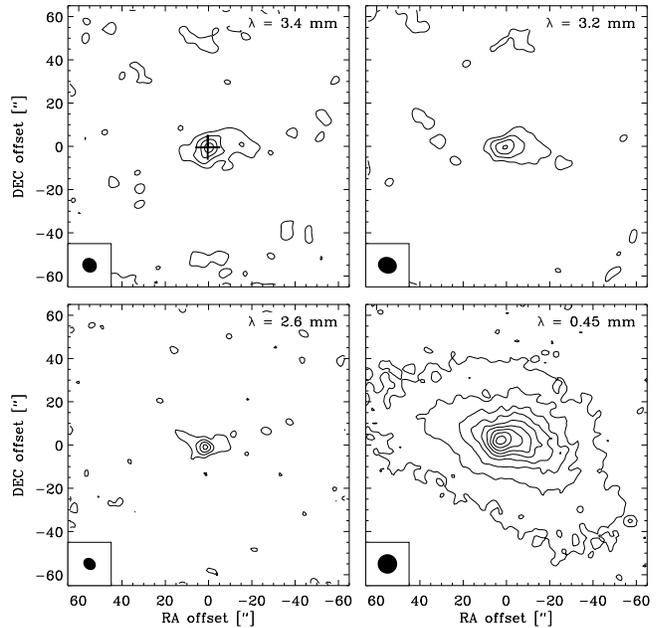}}
\caption{Continuum maps at $\lambda$=3.4, 3.2 and 2.6~mm from the OVRO
observations compared to the SCUBA 450~$\mu$m map. The contours are in
levels of 2$\sigma$ as given in Table~\ref{uvfits}. The position of
the radio source from \cite{beltran01} is indicated by the black '+'
in the upper left, $\lambda= 3.4$~mm, map.}\label{continuummaps}
\end{figure}

\begin{figure*}
\resizebox{\hsize}{!}{\includegraphics{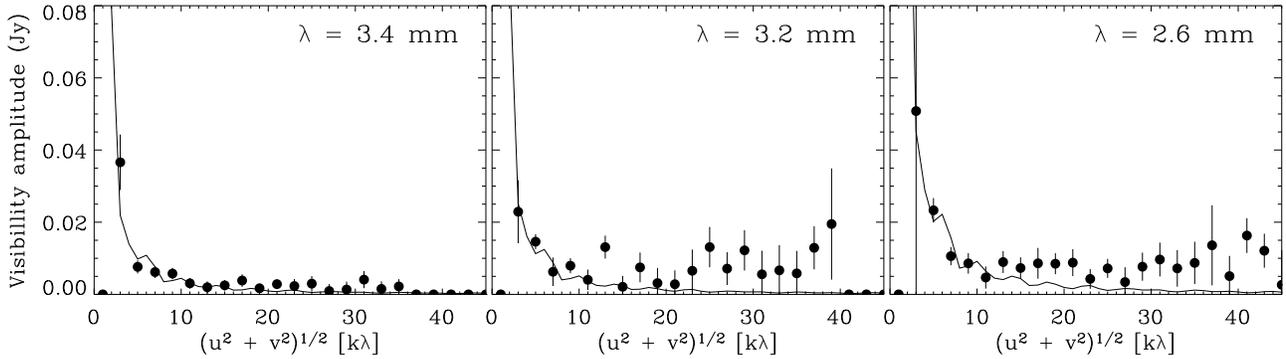}}
\caption{Plots of visibility amplitudes vs. projected baseline line
length for the three continuum datasets at $\lambda$=3.4, 3.2 and
2.6~mm. The solid circles indicate the observations and the lines
the predictions from the model for the L483
envelope.}\label{contuvamp}
\end{figure*}

The continuum maps clearly show the central protostar close to the
phase center (Fig.~\ref{continuummaps}). The emission appears resolved
and roughly follows the morphology from the larger scale SCUBA
maps. Results of Gaussian fits to the visibility data are given in
Table~\ref{uvfits}. The envelope model of \cite{jorgensen02} predicts
total integrated fluxes ranging from 0.16 to 0.32~Jy at wavelengths
from 3.4 to 2.6~mm. Fig.~\ref{contuvamp} shows that the envelope model
reproduces the observed fluxes on all the baselines, but the total
flux recovered in the interferometer maps is only 5-10\% of the total
model flux. This is likely due to the lack of sensitivity in the
interferometer observations to extended emission.

\begin{figure}
\resizebox{\hsize}{!}{\includegraphics{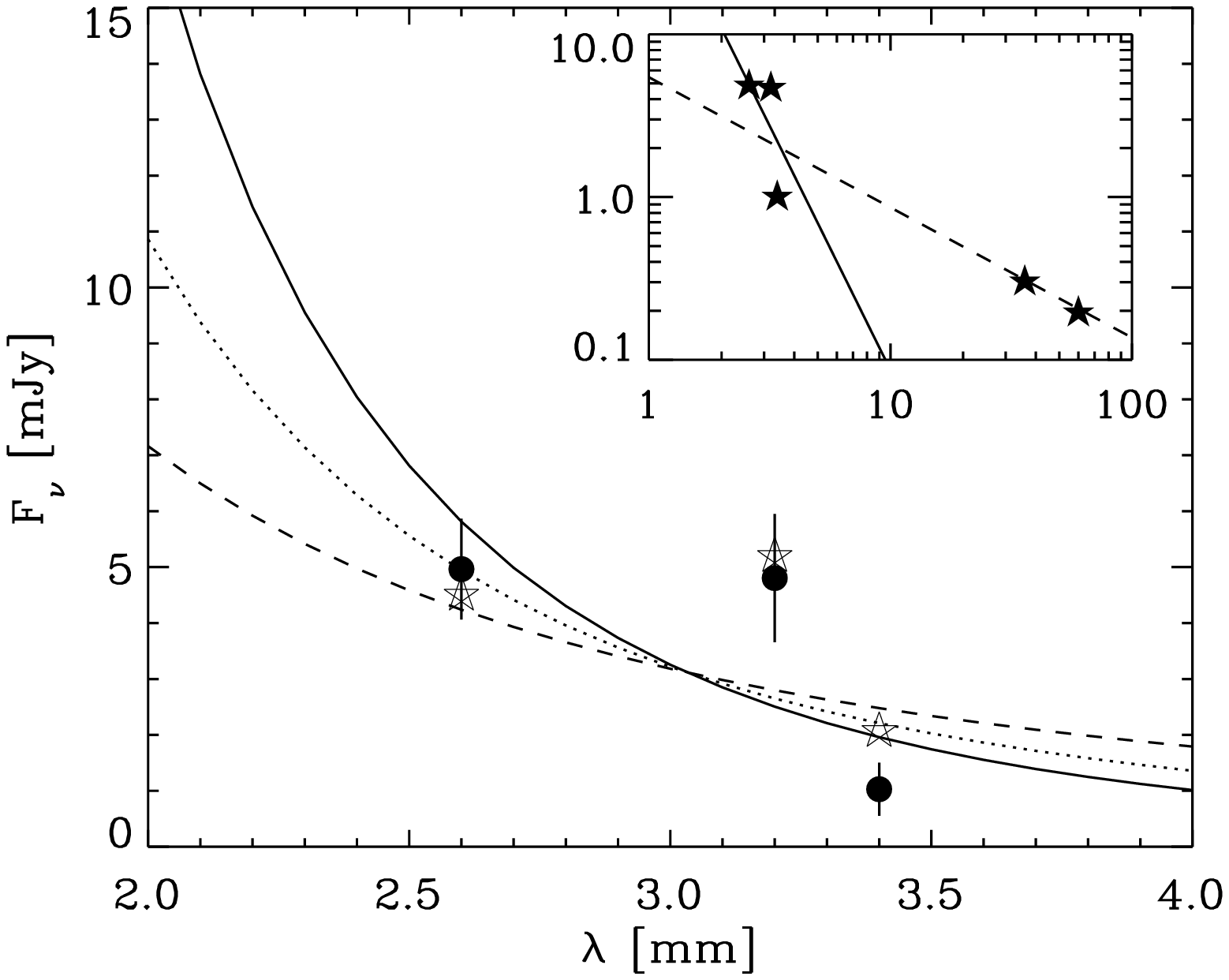}}
\caption{Compact (unresolved) emission as a function of wavelength at
  2.6, 3.2 and 3.4~mm. For each data point the expected flux in the
  absence of any point source is indicated by an open
  star. Fitted power-law distributions with spectral indexes of 2, 3
  and 4 are shown with the dashed, dotted and solid lines,
  respectively. The insert show the millimeter fluxes compared to the
  centimeter data from \cite{beltran01}. Here the dashed line
  indicates the flux from centimeter wavelength with a spectral index
  of 0.8. The solid line indicates a spectral index of 3,
  corresponding to optically thin emission from dust with
    opacities $\kappa_\nu \propto \nu^\beta$ where $\beta=1$ at
  millimeter wavelengths.}\label{spectralindex}
\end{figure}

From Fig.~\ref{contuvamp} it is also evident that very little emission
is picked up on baselines longer than 20-25~k$\lambda$ (i.e., at
scales less than 10\arcsec). Fig.~\ref{spectralindex} shows the fitted
point source fluxes after subtracting the envelope model ($F_{\rm
compact}$, listed in Table~\ref{uvfits}). It can be seen that the
emission is consistent with the zero-expectation level (the expected
amplitude signal due to noise alone in the absence of source emission)
within 1$\sigma$ (3.2 and 3.4~mm observations) and 2$\sigma$
(2.6~mm). This contrasts the situation for \object{NGC~1333-IRAS2A}
\citep{n1333i2art} and \object{IRAS~16293-2422} and \object{L1448-C}
\citep{hotcorepaper} which show compact emission not explained by the
envelope models. In fact the upper limit to the compact flux at 2.6~mm
restricts the mass of a possible disk to $\lesssim$~0.04~$M_\odot$
assuming a temperature of 30~K and optically thin emission. Radio
measurements of L483 at centimeter wavelengths \citep{beltran01} show
a central source of 0.20~mJy (6~cm) and 0.31~mJy (3.6~cm). These radio
observations are compared to the millimeter data from this paper in
the insert in Fig.~\ref{spectralindex}. As can be seen, a spectral
index consistent with optically thin emission cannot simultaneous
explain the millimeter and centimeter observations; however, a flat,
positive spectral index would be consistent with the emission at
centimeter wavelengths and the limits on the millimeter
observations. This would be the case if the flux from centimeter to
millimeter wavelengths is from thermal free-free emission as suggested
by \cite{beltran01}.

\begin{table*}
\caption{Results from fitting Gaussians to the visibility data for the
three continuum datasets.}\label{uvfits}
\begin{center}
\begin{tabular}{llll}\hline\hline
Wavelength [mm]           & 3.4                &  3.2               & 2.6 \\ \hline
RMS [mJy beam$^{-1}$]     & 0.5                &  0.9               & 0.8             \\
Beam size (HPBW) [\arcsec]            & 7.1$\times$6.5     & 8.8$\times$7.6     & 6.2$\times$5.2  \\
$F_{\rm total}$ [mJy]     & 13$\pm 2$          & 12$\pm 2$          & 11$\pm 2$       \\
X-offset [\arcsec]        & --0.5$\pm 0.9$     & \phantom{--}0.1$\pm 0.8$    &  \phantom{--}1.5$\pm 0.3$    \\
Y-offset [\arcsec]        & --0.9$\pm 0.7$     & --0.3$\pm 0.5$     & --0.9$\pm 0.3$  \\ 
$F_{\rm compact}$$^{a}$ [mJy]   & $1.0\pm 0.5$ (2.0) & $4.8\pm 1.0$ (5.2) & $5.0\pm 0.9$ (4.5) \\ \hline
\end{tabular}
\end{center}

$^{a}$Fitted emission from a compact source after subtraction of the
model prediction for the envelope emission. The numbers in parentheses
indicate the zero-expectation level, i.e., the expected amplitude
signal due to noise alone in the absence of any source.
\end{table*}

\section{Line emission}\label{lineobs}
\subsection{Morphology}\label{sdcompare}
\begin{figure*}
\resizebox{\hsize}{!}{\includegraphics{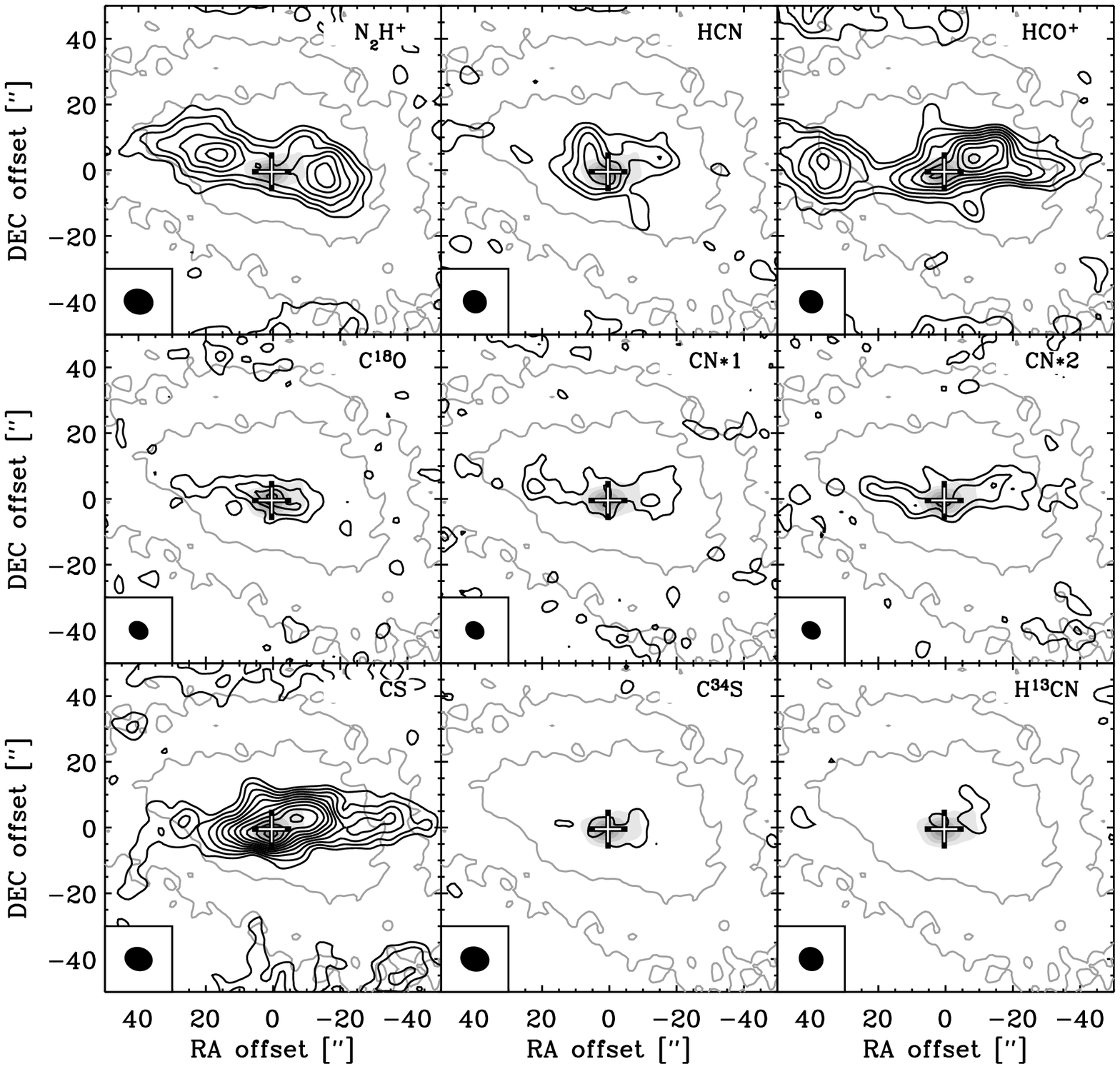}}
\caption{Overview of all observed lines at OVRO. The black line
contours indicate the line emission integrated from 2.5 to 8.0~km~s$^{-1}$
for CS, HCN, HCO$^+$, and N$_2$H$^+$ and from 4.0 to 6.5~km~s$^{-1}$ for the
remaining species with levels given in steps of 2$\sigma$. The grey
line contours indicate the SCUBA 450~$\mu$m emission, the grey-scale
map indicates the 3.4~mm continuum emission. For HCN, H$^{13}$CN and
N$_2$H$^+$ the emission is integrated over the main hyperfine
component. The CN$\ast 1$ and CN$\ast 2$ maps show the emission of the
two CN 1--0 hyperfine components at 113.489 and 113.491~GHz,
respectively.}\label{momentone}
\end{figure*}

Maps of the total integrated emission of each of the observed lines
are shown in Fig.~\ref{momentone}. All species are detected toward the
continuum position and most are stretched in the east-west direction
of the elongated core from the SCUBA maps.

The N$_2$H$^+$ emission is seen to follow the contours of the dust
emission from the SCUBA continuum maps closely, except toward the
position of the central star where the emission appears
suppressed. This gives N$_2$H$^+$ emission a unique ``peanut-shape''
compared to the remaining species, with the strongest emission at two
lobes, respectively east and southwest of the continuum peak
position. C$^{18}$O in contrast is mainly seen very closely confined to
the region around the central protostar. The emission is resolved and
interestingly shows an almost triangular shape similar to that found
in the SCUBA maps (e.g., lower right panel of
Fig.~\ref{continuummaps}): the two shortest sides in the north-south
direction on the western side of the core and in the east-west
direction on the northern side of the core, and the longest side
stretching from the northeast to the southwest.

\begin{figure}
\resizebox{\hsize}{!}{\includegraphics{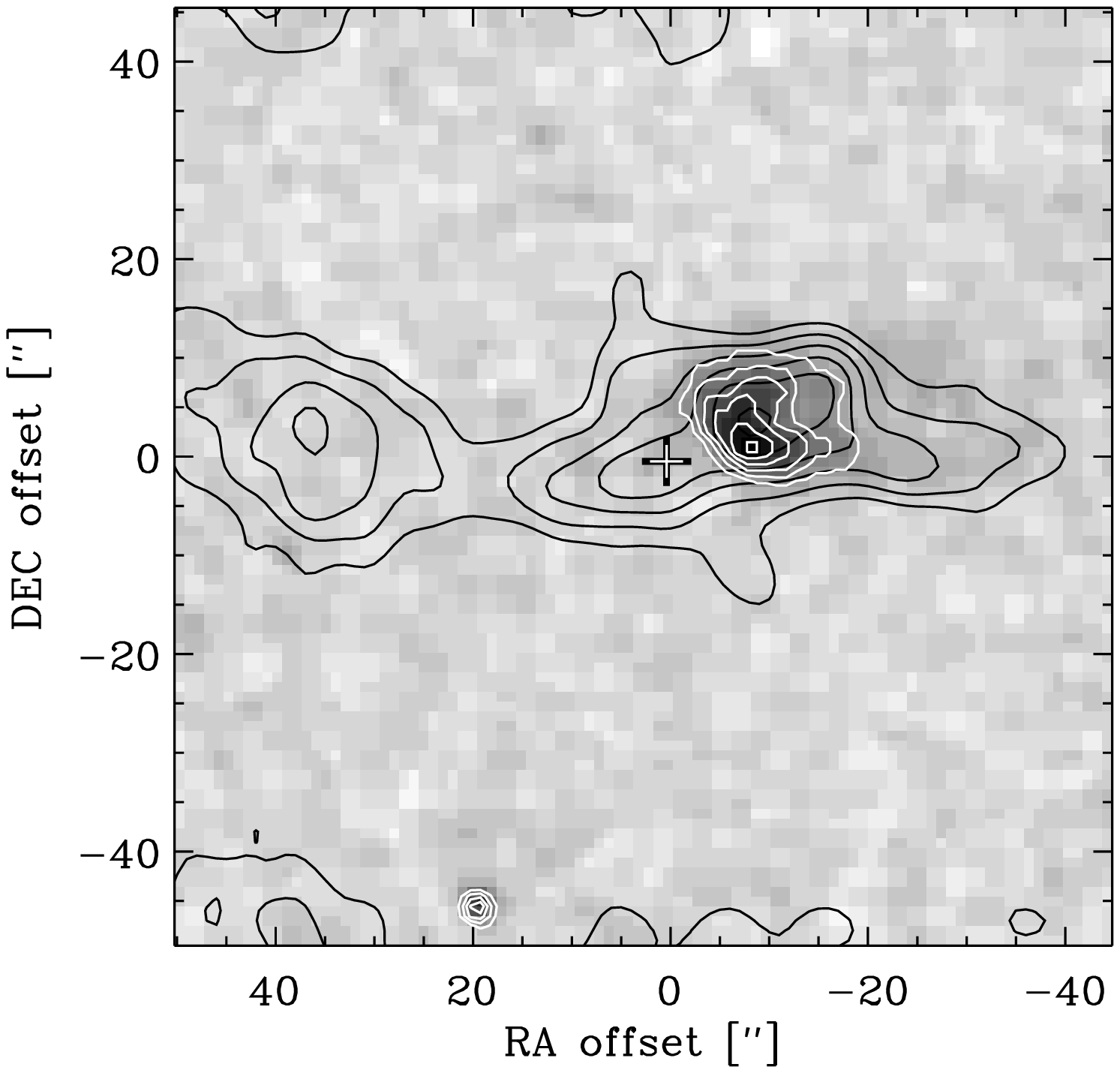}}
\caption{2MASS near-infrared $K_{\rm s}$ image of L483 (grey-scale and
white contours) compared to the integrated HCO$^+$ 1--0 emission
(black contours; in steps of 3$\sigma$). Notice the near-perfect
agreement between the infrared nebulosity and the outflow emission
probed by HCO$^+$.}\label{tomass}
\end{figure}

HCO$^+$ shows the most extended emission. It has a slightly curved
structure around the central protostar, being almost anti-correlated
with the N$_2$H$^+$ emission. A similar ``twist'' can be seen in the
single-dish CO maps of \cite{hatchell99} and \cite{tafalla00} and in
general the HCO$^+$ appears closely correlated with the outflow, as
was also concluded based on the BIMA HCO$^+$ maps by
\cite{park00}. Fig.~\ref{tomass} compares the \emph{2MASS All-Sky
Quicklook}\footnote{The Two Micron All Sky Survey (2MASS) is a joint
project of the University of Massachusetts and the Infrared Processing
and Analysis Center/California Institute of Technology, funded by the
National Aeronautics and Space Administration and the National Science
Foundation. The presented image was obtained from the NASA/IPAC
Infrared Science Archive, which is operated by the Jet Propulsion
Laboratory, California Institute of Technology, under contract with
the National Aeronautics and Space Administration.} $K_{\rm s}$ image
of L483 with the observed HCO$^+$ emission. Near-perfect overlap is
seen between the infrared and HCO$^+$ emission in the western outflow
lobe. This in agreement with the suggestion by \cite{fuller95} that
the infrared emission is light from the central protostar scattered
off the outflow cavity walls. The HCN and CS emission appear to be
more centrally concentrated around the central protostar but do show
traces of the same elongated east-west feature. In addition a
north-south feature perpendicular to the main elongation is seen in
both species extending over $\approx$ 20\arcsec. Their weaker isotopic
species, C$^{34}$S and H$^{13}$CN, only show unresolved emission around the
central continuum peak.

CN shows a very narrow feature elongated in the east-west
direction. Toward both N$_2$H$^+$ lobes the CN emission makes a twist as
seen in the HCO$^+$ maps, but for CN the emission is much narrower. In
fact CN is found to trace the boundary between the N$_2$H$^+$ and
HCO$^+$ emission (see Sect.~\ref{cnchem}). The two CN hyperfine
components otherwise show similar structures indicating quite
homogeneous excitation conditions along the main elongation.

\begin{figure*}
\resizebox{\hsize}{!}{\includegraphics{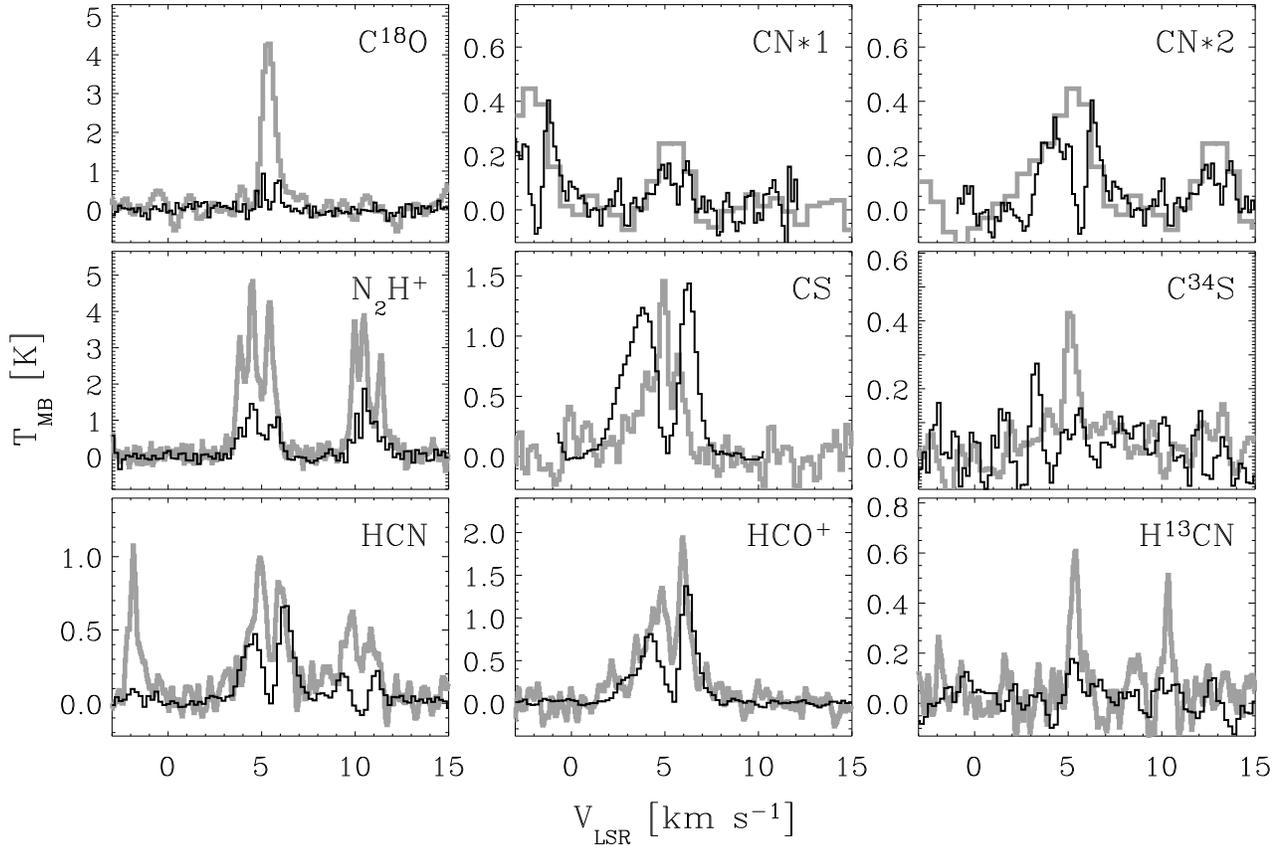}}
\caption{Comparison between the single-dish spectra (grey) and spectra
extracted from the interferometer datacubes convolved with the
single-dish beam (black). Each spectrum from the interferometer data
has furthermore been multiplied by a factor 5 to make comparison
between the lineshapes easier. As in Fig.~\ref{momentone} the
spectra marked CN$\ast 1$ and CN$\ast 2$ shows the emission of the two
CN 1--0 hyperfine components at 113.489 and 113.491~GHz,
respectively.}\label{resolve_out}
\end{figure*}

As found for the continuum observations, the interferometer picks up
only a fraction of the line emission seen by the single-dish
telescope. Fig.~\ref{resolve_out} compares the single-dish spectra
from the Onsala 20~m telescope \citep{jorgensen02,paperii} with
those from the interferometer datacubes convolved with Gaussians
similar to the size of the single-dish beam. The interferometer
spectra recover only a fraction of the emission close to the systemic
velocity of the cloud ($\approx 5.3$~km~s$^{-1}$) whereas that in the wings
agrees better, i.e., is less subject to resolving out.

The large degree of resolving out, together with the asymmetry of the
source, naturally complicates the interpretation of the line
observations and, in particular, renders direct calculations of, e.g.,
column densities based on the absolute values of the interferometer
observations impossible. The interferometer observations do, however,
give the location and sizes of the brightest emission and one can
thereby utilize comparisons between the maps of the different species
to address the spatial variations of the chemistry. Also, predictions
from the line radiative transfer model can be used for comparison to
the interferometer observations, especially for species that show a
relatively simple structure such as C$^{18}$O (see discussion in
Sect.~\ref{cochem}).

\subsection{Velocity field}
\begin{figure*}
\resizebox{\hsize}{!}{\includegraphics{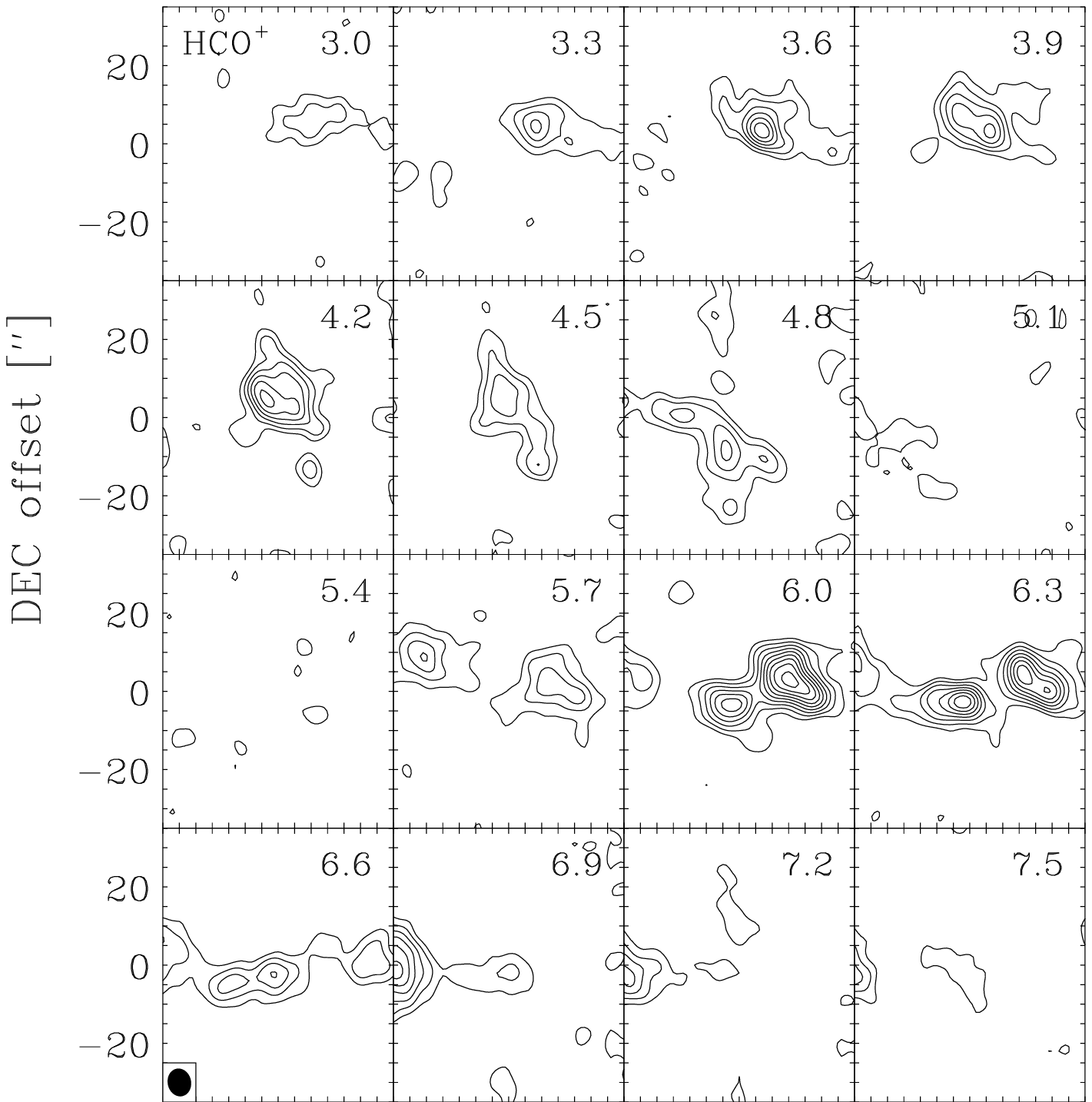}\includegraphics{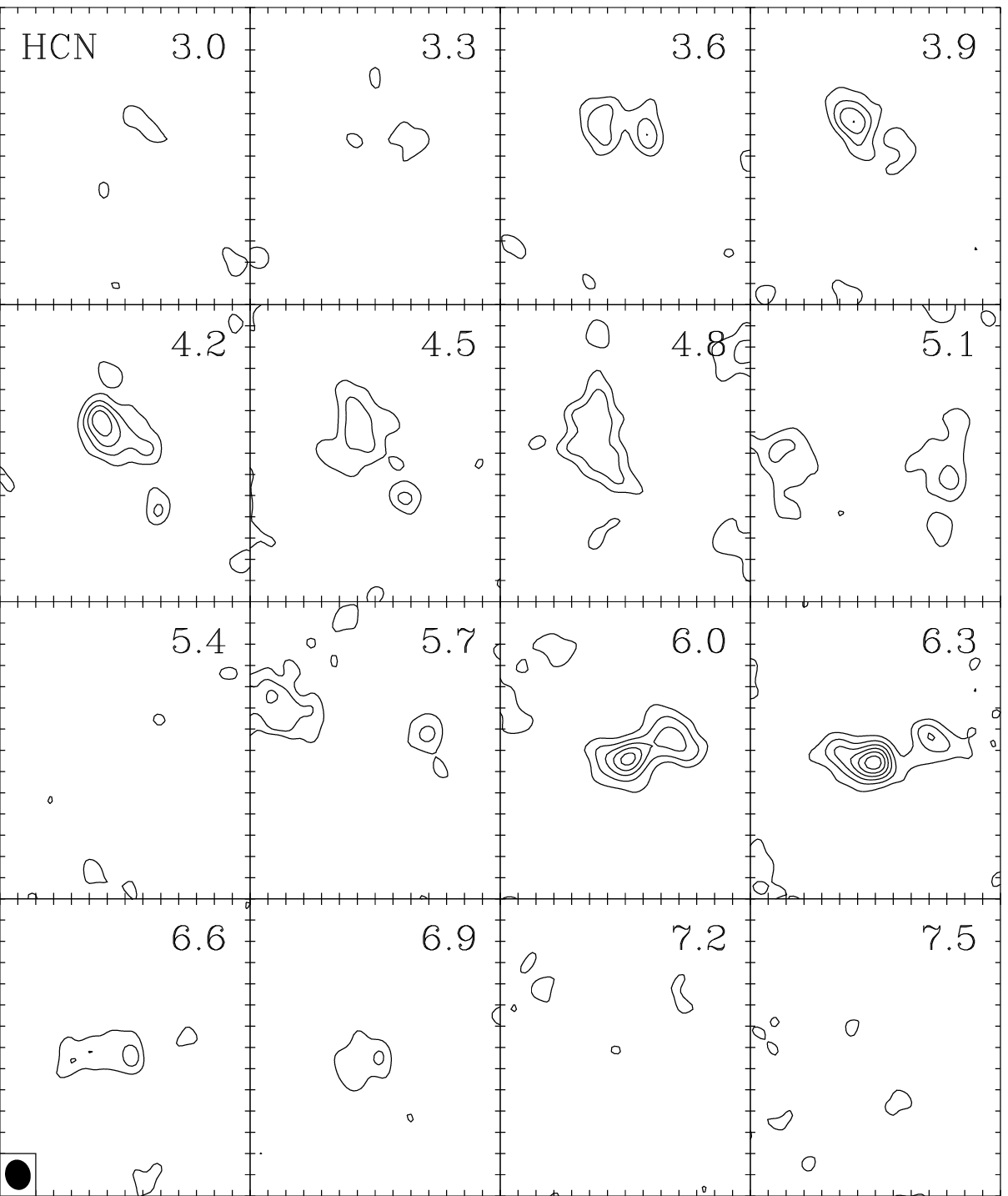}}
\resizebox{\hsize}{!}{\includegraphics{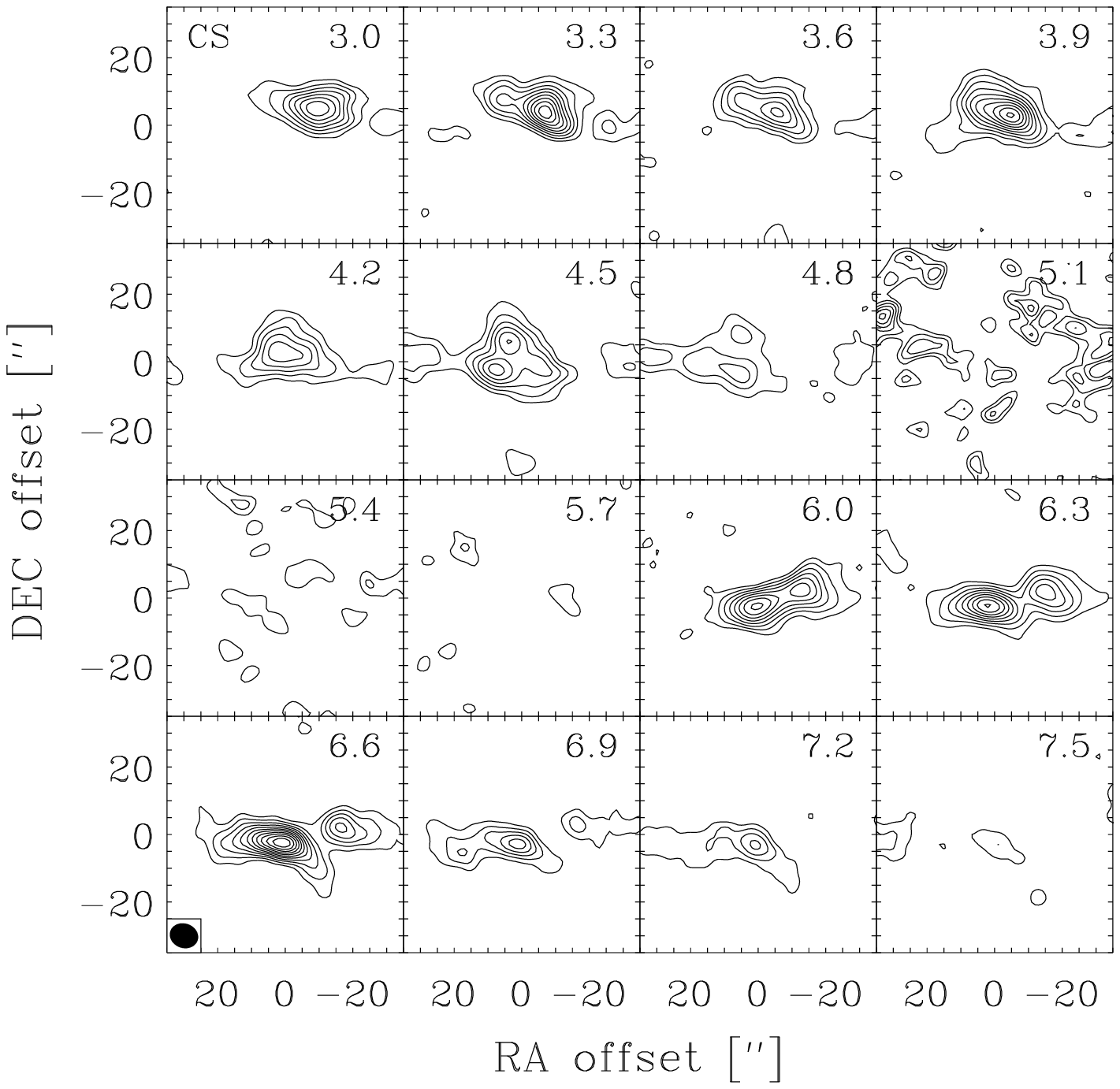}\includegraphics{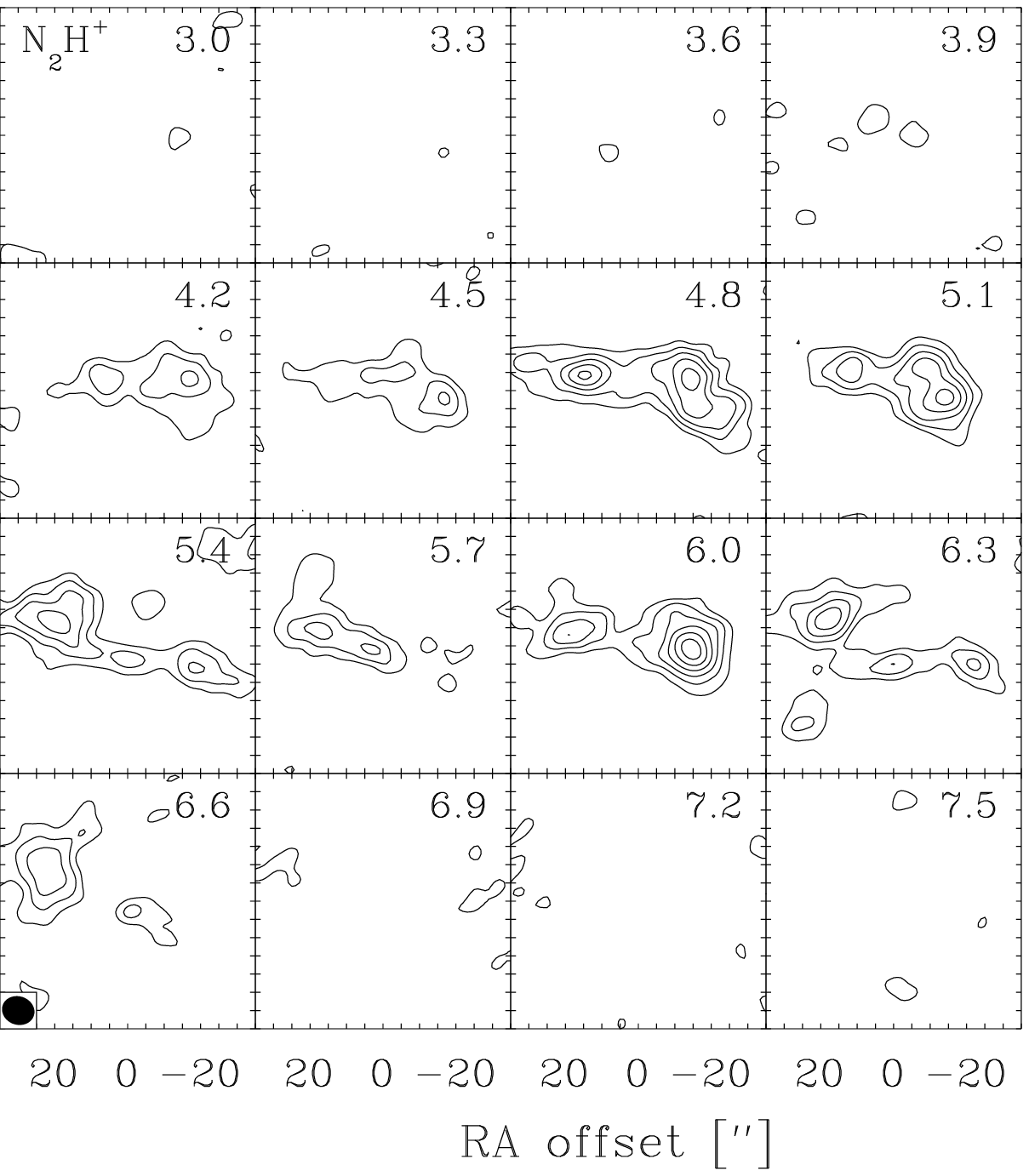}}
\caption{Channel maps for the HCO$^+$, HCN, CS and N$_2$H$^+$
emission. Contours are given in steps of 3$\sigma$.}\label{channel}
\end{figure*}

Velocity gradients are seen for a number of species toward L483 as
illustrated in channel maps of the lines of HCO$^+$, HCN, CS and
N$_2$H$^+$ in Fig.~\ref{channel}. The HCO$^+$ maps clearly show the twist
of the outflow toward the east for the red-shifted lobe, with distinct
cores at offsets (15\arcsec,-5\arcsec) at 6.6~km~s$^{-1}$ and the edge of
the channel maps at $>$~6.6~km~s$^{-1}$. The first of these two clumps are
also clearly seen in the CS channel maps at the same velocity. This
core is merging with the central protostellar core at offsets
(3\arcsec,-3\arcsec) seen toward all four species. In the maps of the
blue-shifted emission the northern part of the central core is also
seen at (0\arcsec,5\arcsec) for velocities around 4.2~km~s$^{-1}$. The
outflow emission is less obvious in the blue-shifted lobe. A last
feature is a red-shifted core at (-15\arcsec,0\arcsec) in HCN, HCO$^+$
and CS at velocities of 6.3~km~s$^{-1}$. The two peaks of the N$_2$H$^+$ emission
are seen to be unrelated to emission in the other maps and peak at
respectively 5.1 and 6.3~km~s$^{-1}$.

As discussed by \cite{n1333i2art}, it may be somewhat problematic
directly inferring velocity gradients from interferometer maps with
missing short-spacings since the interferometer is predominantly
sensitive to material at the more extreme velocities. As can be seen
from Fig.~\ref{resolve_out}, the emission in the line wings is indeed
least subject to resolving out compared to the emission close to the
systemic velocity, but in general the observed line profiles appear
similar from both single-dish and interferometer observations
indicating that they are still probing much the same material.

\begin{figure}
\resizebox{\hsize}{!}{\includegraphics{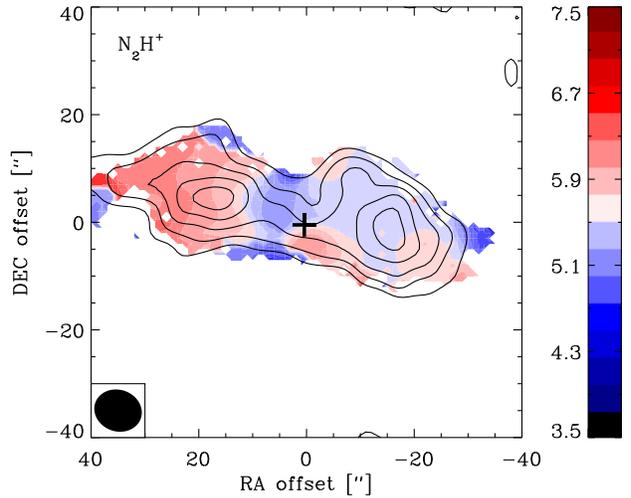}}
\caption{The velocity field from the N$_2$H$^+$ maps as colored
contours compared to the integrated emission (black line contours).
The velocity has been derived by fitting the 7 hyperfine components of
the N$_2$H$^+$ 1--0 line over the entire map.}\label{n2hp_velfield}
\end{figure}

For N$_2$H$^+$, the hyperfine splitting further complicates the
interpretation of the velocity field. As was also found for
\object{NGC~1333-IRAS2A}, however, each component is rather narrow, not
affected by the outflow, and well-represented by single Gaussians. To
derive the velocity field probed by N$_2$H$^+$, the emission from the 7
hyperfine components was therefore fit simultaneously to the spectra
in each pixel. The linewidths for the individual hyperfine components
were taken to be constant and fit together with the systemic velocity
field, the overall normalization of the entire hyperfine group and the
relative intensities of the hyperfine components. The fitted velocity
field is illustrated in Fig.~\ref{n2hp_velfield}. An overall red-blue
asymmetry is found between the eastern and western lobes. Toward the
source position it is seen that the velocity gradient is directed
toward the north-south as in the cases of HCN and CS.
The very characteristic two lobes in N$_2$H$^+$ are also seen in NH$_3$ by
\cite{fuller00}, who furthermore found that the NH$_3$ emission shows
a two velocity component structure with the NE lobe red-shifted to
5.5-5.6~km~s$^{-1}$ and the SW lobe blue-shifted to 5.3-5.4~km~s$^{-1}$. This is
exactly what is found for our N$_2$H$^+$ data, indicating that the NH$_3$
and N$_2$H$^+$ emission probe the same material. Both \cite{fuller95}
and \cite{tafalla00} proposed L483 as a core being broken up by the
action of the outflow resulting, for example, in the bipolar
near-infrared nebulosity. The velocity field seen in N$_2$H$^+$ and
NH$_3$ and its similarities with the outflow velocity pattern would
suggest that such an interaction in fact is ongoing although these
species are not directly probing the outflowing gas.

\cite{fuller00} found the most blue-shifted emission toward the
central star with velocities of 5.1-5.3~km~s$^{-1}$, and suggested that the
more blue-shifted emission toward the central star is a result of
infall. In our N$_2$H$^+$ maps it is also seen that the emission has
its most blue-shifted peak about 5\arcsec\ northwest of the
protostellar mm/cm source. However, an accompanying red-shifted
component is present on the other side of the central source (i.e., to
the southeast). This is further supported by our HCN and CS maps which
show a similar velocity pattern around the central protostar, also
seen by \cite{park00} in C$_3$H$_2$ maps. A tempting suggestion for
this velocity gradient is rotation around the central protostellar
object and the propagation axis of the outflow. The CS and HCN lines
have rather high critical densities and appear to probe only the
densest gas on small scales close to the central protostar. Likewise
C$_3$H$_2$ is typically only found in the dense envelope material not
affected by the outflow \citep[e.g.,][]{bachiller97,tafalla97}.

\begin{figure}
\resizebox{\hsize}{!}{\includegraphics{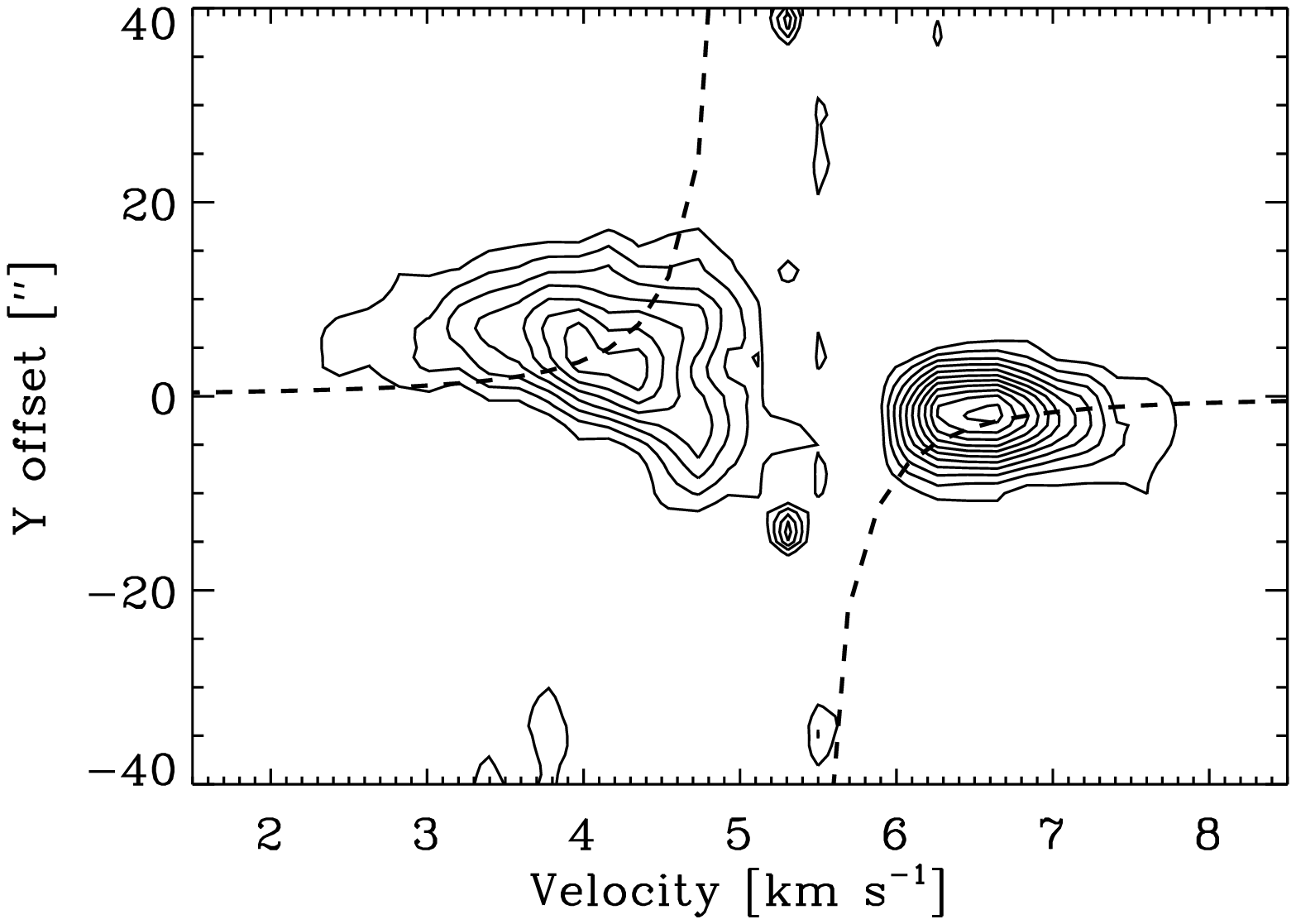}}
\resizebox{\hsize}{!}{\includegraphics{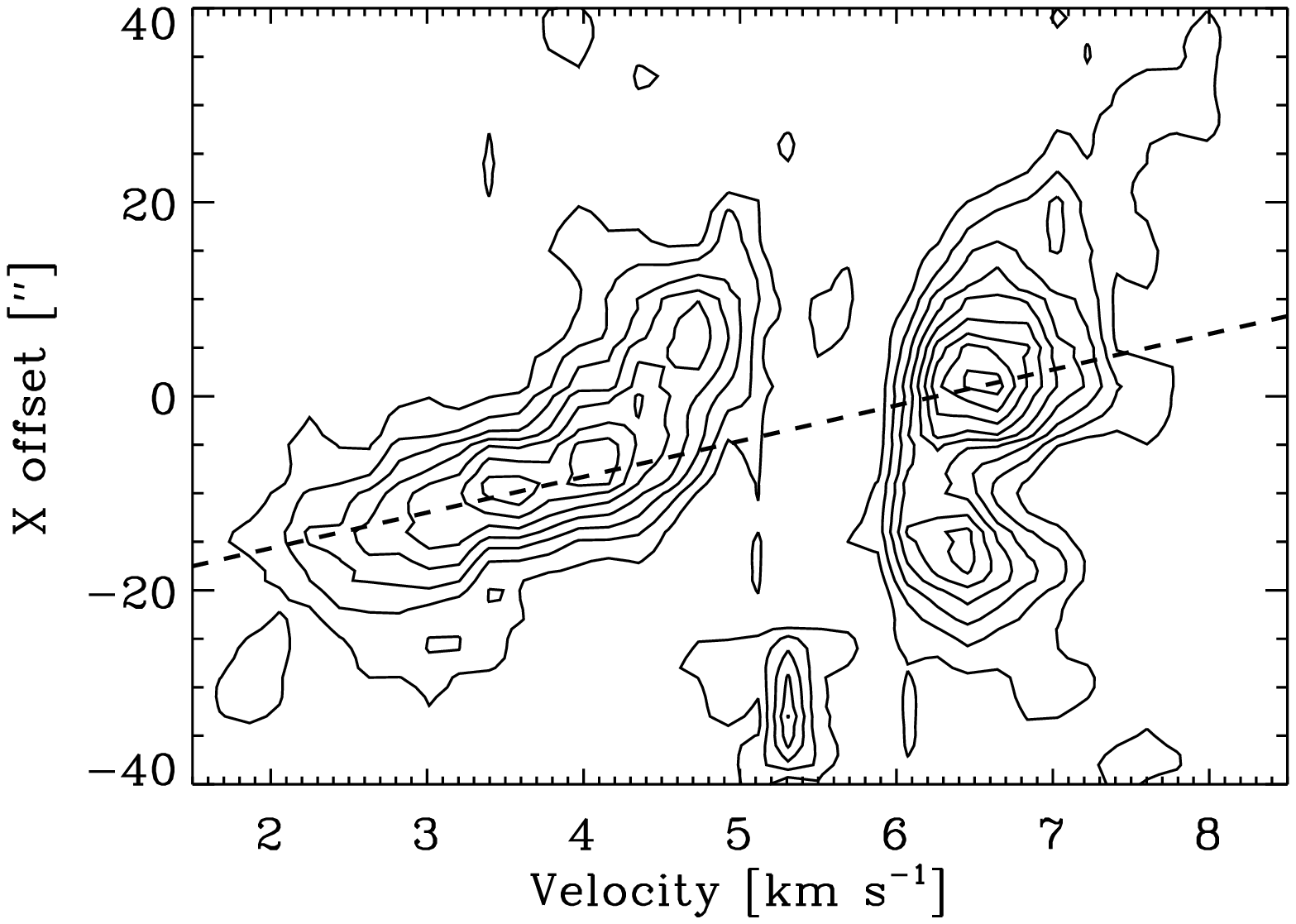}}
\caption{Position-velocity diagram for the CS emission rotated
12$^\circ$, so the Y-axis is in the direction of the velocity gradient
in the dense material close to the central protostar (see also
Fig.~\ref{nthp_co}). The dashed lines indicate Keplerian rotation
around a 1.2~$M_\odot$ central object in the upper panel and linear
expansion of 0.27~km~s$^{-1}$~arcsec$^{-1}$ in the lower panel.}\label{pv_cs}
\end{figure}

Fig.~\ref{pv_cs} shows the position-velocity diagram for the CS
emission. The coordinate system has been rotated to have the Y-axis in
the direction of the north-south velocity gradient around the central
protostar (i.e., 12$^\circ$ from north through east) and the
X-axis in the direction of the outflow perpendicular to this. The
largest velocity gradient is seen in the north-south direction of the
CO outflow. At velocities $\sim$~1~km~s$^{-1}$ from the systemic velocity
the gradient is found to be consistent with linear expansion of
0.27~km~s$^{-1}$~arcsec$^{-1}$ or 2.6~$\times 10^{-4}$~km~s$^{-1}$~pc$^{-1}$ at the distance
(200~pc) of L483.

A clear velocity gradient is also observed in the north-south
direction: plotted on top of the position-velocity diagram is the
predicted rotation curve for Keplerian rotation around a 1.2~$M_\odot$
central object. This velocity gradient fits the data at large
velocities compared to the systemic velocity, but at velocities of
4.5-5~km~s$^{-1}$ the emission appears to stretch southwards in contrast to
the prediction for Keplerian rotation. The cause of this is the offset
in the north-south direction between the red and blue-shifted lobes of
the outflow. Alternative explanations to the pure Keplerian rotation
\citep[e.g., a combination of rotation and infall such as suggested
for another embedded low-mass YSO, TMC1, by][]{hogerheijde01} cannot
be ruled out based on the data presented in this paper, since the
confusion with the outflow around L483 complicates the interpretation.

\section{Discussion}\label{discussion}
\subsection{Thermal structure, depletion of CO and resulting chemistry}\label{cochem}
\begin{figure}
\resizebox{\hsize}{!}{\includegraphics{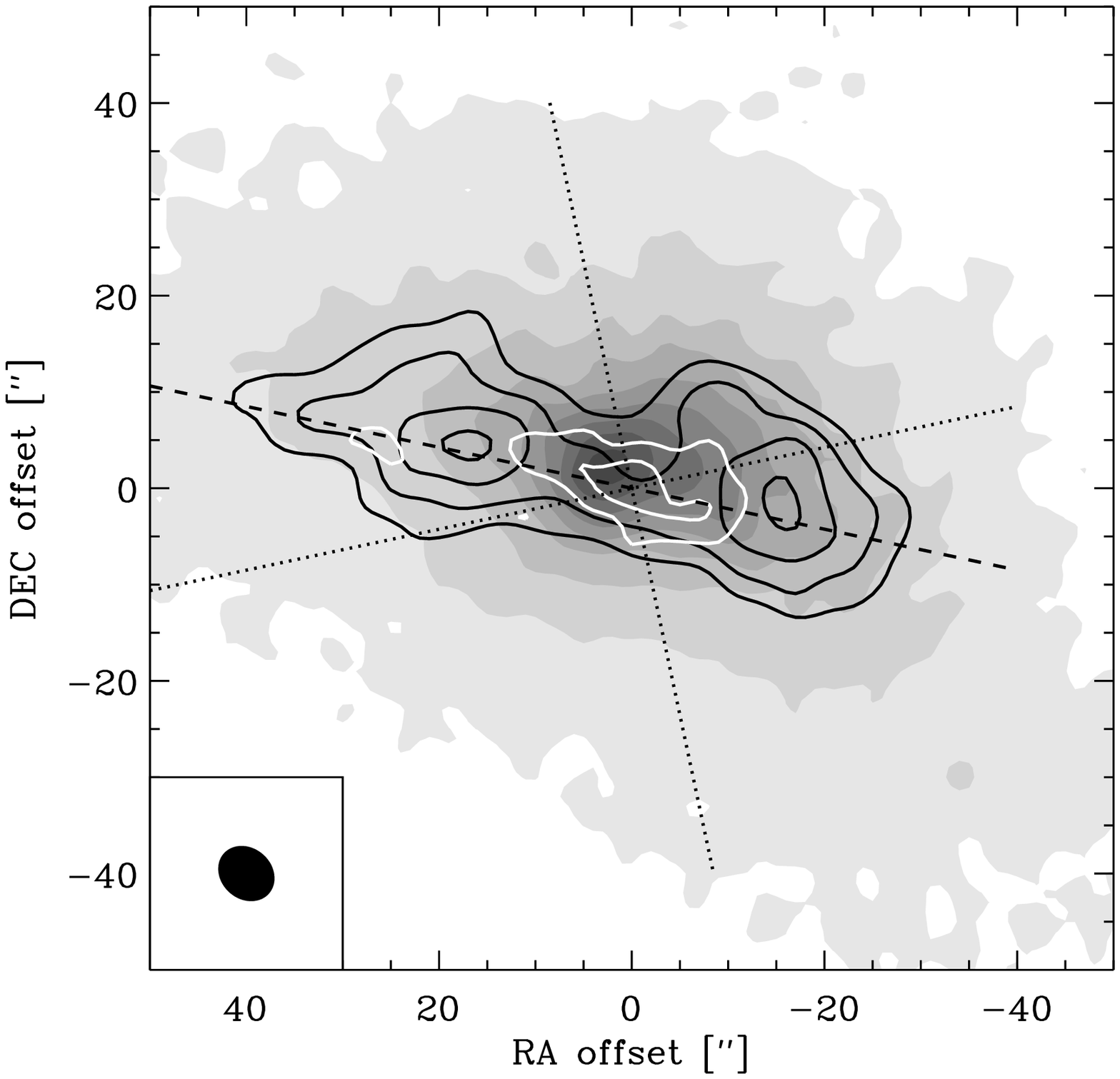}}
\resizebox{\hsize}{!}{\includegraphics{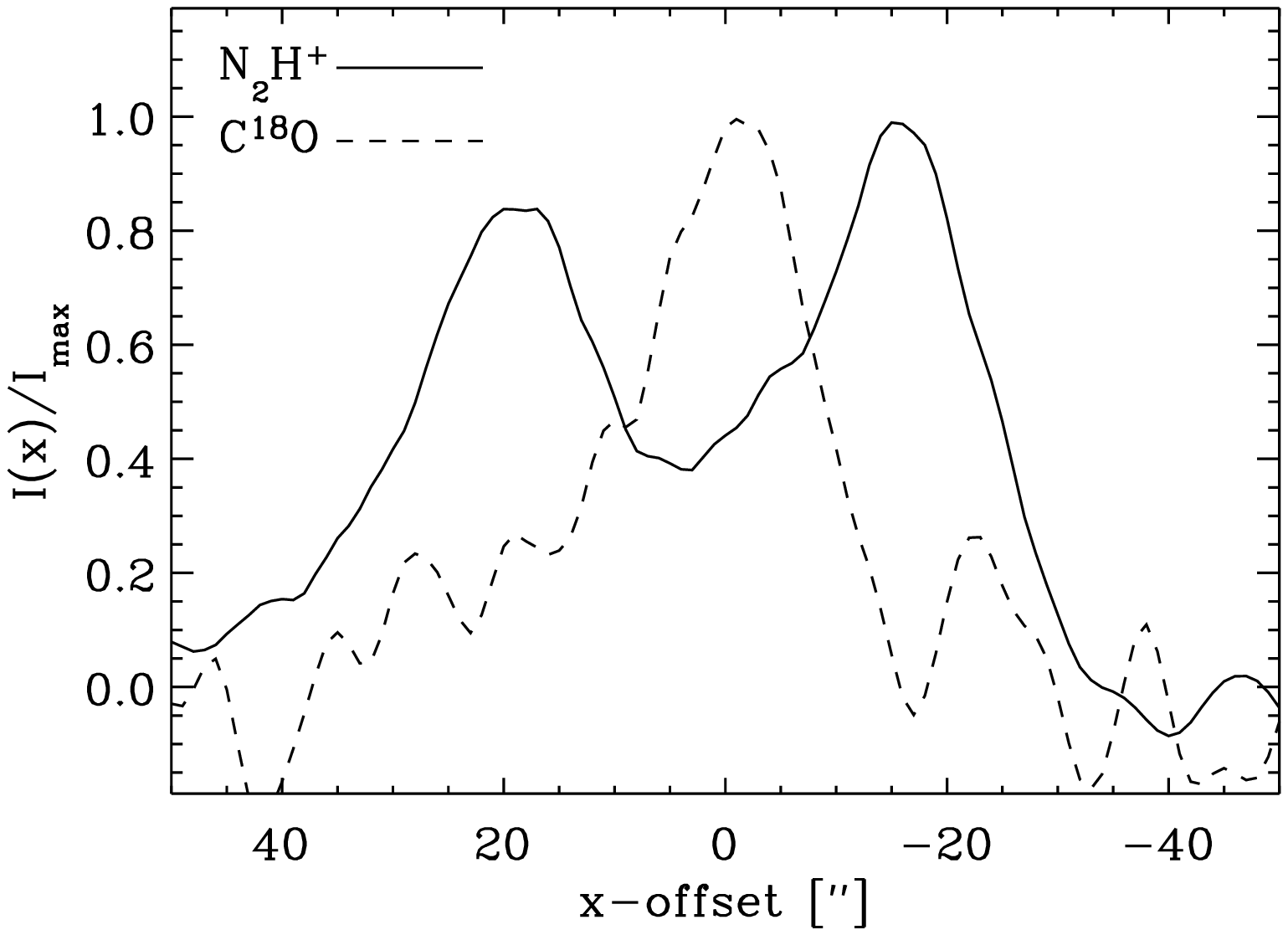}}
\caption{Comparison between N$_2$H$^+$ vs. C$^{18}$O emission. \textit{Upper
panel:} The integrated N$_2$H$^+$ and C$^{18}$O emission (black and white
contours) plotted over SCUBA 450~$\mu$m image (grey-scale). The
flattened direction of the core (i.e., 12$^\circ$ from the east-west
axis) has been indicated by the dashed line. The coordinate system
with the rotation velocity gradient (Fig.~\ref{pv_cs}) has been
indicated by the dotted lines. \textit{Lower panel:} the N$_2$H$^+$ and
C$^{18}$O emission in a strip in the flattened direction of the core
(dashed line in upper panel). Note the clear anti-correlation between
the peaks in C$^{18}$O and N$_2$H$^+$.}\label{nthp_co}
\end{figure}

The most striking feature in the maps of the integrated emission in
Fig.~\ref{momentone} is the characteristic ``peanut-shaped'' N$_2$H$^+$
maps, with N$_2$H$^+$ being absent from the gas-phase close to the central
protostar. This likely reflects destruction of N$_2$H$^+$ by reactions
with CO, as suggested to be the case in protostellar environments by,
e.g., \cite{bergin01} and \cite{n1333i2art,paperii}. \cite{paperii}
found a clear anti-correlation between the abundances of CO and N$_2$H$^+$,
which was suggested to be caused by CO depletion, and thereby less
N$_2$H$^+$ destruction, in the cold part of the envelope. It is therefore
interesting to compare the N$_2$H$^+$ and CO maps presented in this paper:
as further illustrated in Fig.~\ref{nthp_co}, the N$_2$H$^+$ and C$^{18}$O
peaks are clearly anti-correlated with CO being present towards the
central protostar. This supports the claim that the C$^{18}$O maps peak
toward the regions where CO has evaporated from the grain mantles
whereas N$_2$H$^+$ peaks away from the central continuum position where CO
is frozen-out.

\begin{figure}
\resizebox{\hsize}{!}{\includegraphics{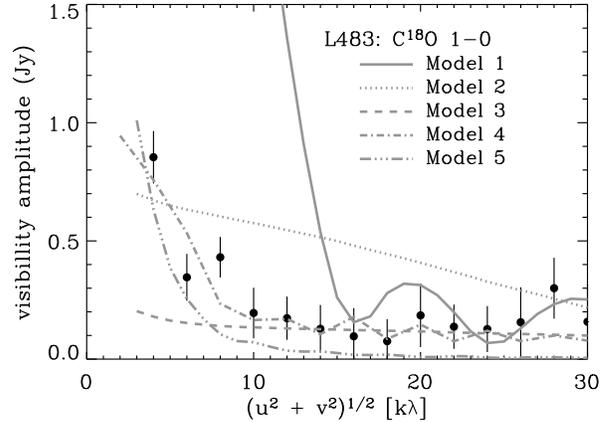}}
\caption{Comparisons between the observed visibilities (solid
circles) and model predictions (grey lines) for the C$^{18}$O emission
averaged over the observed line (4.0 to 6.5~km~s$^{-1}$) plotted
vs. projected baseline length. Model 1-3 are jump models with $T_{\rm
ev}$ of 20, 30 and 40~K, respectively. Model 4 is a drop model with
$T_{\rm ev}=40$~K and $n_{\rm de}=1.5\times 10^5$~cm$^{-3}$. Model 5
is a constant abundance model. Note that only model 4 is consistent
with all single-dish line intensities presented in \cite{jorgensen02}
as shown in Table~\ref{chi2table}.}\label{c18o_uvmodel}
\end{figure}

As discussed in Sect.~\ref{sdcompare}, a significant fraction of the
line emission is resolved out. Since the structure of the C$^{18}$O
emission appears relatively simple, we can calculate the emission from
the envelope adopting the physical structure and CO abundances from
\cite{jorgensen02} as shown in Fig.~\ref{c18o_uvmodel}. This model is
compared to ``jump'' models where the CO abundance increases at radii
where the temperature is above a given threshold mimicking evaporation
of CO. As can be seen from Fig.~\ref{c18o_uvmodel}, the constant
abundance model provides a better fit to the interferometer data: the
``jump'' models either overproduce the observed emission at
intermediate baselines (models with $T_{\rm ev}=20$~K and 30~K) or do
not produce a large increase at the shortest baselines (model with
$T_{\rm ev}=40$~K). As discussed in \cite{jorgensen02,paperii},
however, although the constant CO abundance and jump models can
successfully explain the intensities of the higher $J=2-1$ and $J=3-2$
CO lines, they generally under-produce the low $J=1-0$ emission. It is
therefore not unexpected that such models have problems explaining the
radial variation of the observed CO emission.

As an alternative, \cite{coevollet,paperii} suggest a ``drop'' model
where CO is frozen out only in regions where the temperature is lower
than the evaporation temperature $T_{\rm ev}$, while at the same time
the density is higher than a given density, $n_{\rm de}$, so that the
depletion timescale is shorter than the age of the core. This is
similar to the case seen for pre-stellar cores
\citep[e.g.,][]{caselli99,tafalla02,lee03} where CO is frozen out
towards the core center where the density high, and the corresponding
timescale for depletion thereby shorter than the lifetime of the
core. Such ``drop'' models can explain both high and low $J$ CO lines,
and the difference in abundances between more and less massive
envelopes seen in \cite{jorgensen02}, and they work very well in
describing the radial variations of molecules such as H$_2$CO
\citep{hotcorepaper}.

Models for the single-dish CO emission were calculated for L483 and
compared to the line intensities reported in \cite{jorgensen02}. A
drop model with $n_{\rm de} = 1.5\times 10^{5}$~cm$^{-3}$ and $T_{\rm
ev}=40$~K, an undepleted CO abundance of 2.7$\times 10^{-4}$ and depletion in
the drop region by a factor 50 was found to provide good fits to the
observed single-dish emission as indicated by the reduced $\chi^2$ for
the fit as given in Table~\ref{chi2table}. This model has been
compared to the interferometer data in Fig.~\ref{c18o_uvmodel} and is
found to give a good fit to the observed C$^{18}$O 1--0 emission at
practically all baselines. Freeze-out up to a temperature of 40~K is
needed in order not to produce too much emission at the intermediate
baselines and the inclusion of an outer low density region, where the
CO abundance increases, nicely reproduces the observed trend at the
shortest baselines.

\begin{table*}
\caption{$\chi^2$ for fit of models to single-dish and interferometer data.}\label{chi2table}
\begin{center}\begin{tabular}{lllll} \hline\hline
Model & & $\chi^2_{\rm red,s}$$^{a}$ & $\chi^2_{\rm red,i}$ $^{b}$            \\ \hline
1: & Jump $T_{\rm ev}$~=~20~K                                          & 221  & 478 \\
2: & Jump $T_{\rm ev}$~=~30~K                                          & 18.9 & 8.0 \\
3: & Jump $T_{\rm ev}$~=~40~K                                          & 10.8 & 4.7 \\
4: & Drop $T_{\rm ev}$~=~40~K; $n_{\rm de}$~=~1.5$\times 10^5$~cm$^{-3}$     & 1.5  & 1.3 \\
5: & Constant                                                          & 8.0  & 3.1 \\ \hline
\end{tabular}
\end{center}

$^{a}$Reduced $\chi^2$ for fits to single-dish line
intensities. $^{b}$Reduced $\chi^2$ for fits to interferometer data as
shown in Fig.~\ref{c18o_uvmodel}.
\end{table*}

\subsubsection{Chemical implications}
The CO abundance structure has important chemical implications. As
described above, distinct zones with and without depletion of CO also
account for the characteristic shape of the N$_2$H$^+$ emission. In
their NH$_3$ maps \cite{fuller00} found a very similar structure with
a ``valley'' devoid of emission in the north-south direction through
the position of the protostar but otherwise following the large scale
dust continuum emission from the SCUBA maps. \cite{caselli02benson}
likewise suggested a close correlation between emission seen in
N$_2$H$^+$ and NH$_3$ based on a large single-dish survey of
N$_2$H$^+$ emission toward dense cores. The precursor for both
N$_2$H$^+$ and NH$_3$ is N$_2$ which due to its low binding energy
does not deplete unless at very high ages and densities
\citep[e.g.][]{bergin97,bergin02}. NH$_3$ maintains a balance with the
overall chemical network of N-bearing species, including N$_2$H$^+$,
predominantly through reactions with NH$_4^+$, and at low (depleted)
CO abundances these species establish a closed network. At
``standard'' CO abundances, reactions between C$^+$ and NH$_3$,
however, drive the nitrogen into H$_2$NC$^+$ and from there to CN, HNC
and HCN, whereas the N$_2$H$^+$ is destroyed through reactions with
CO.

\begin{figure}
\resizebox{\hsize}{!}{\includegraphics{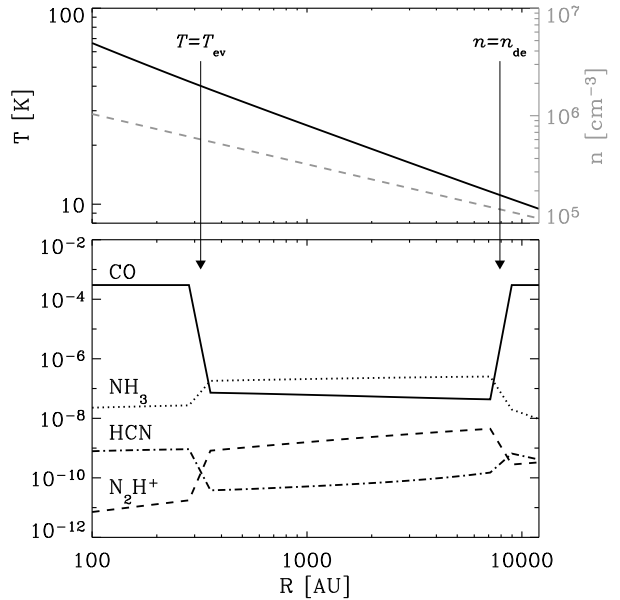}}
\caption{Molecular abundances as function of radius for L483, adopting
the chemical model appropriate for \object{IRAS~16293-2422} at
10$^4$~years \citep{doty04}. \textit{Upper panel:} the density and
temperature profile for L483 shown with dashed and solid lines,
respectively. \textit{Lower panel:} the abundances of CO, NH$_3$,
N$_2$H$^+$ and HCN from the model calculations. The CO abundance
(solid line) structure is assumed to follow a ``drop'' abundance
structure with an evaporation temperature, $T_{\rm ev}$, of 40~K and a
depletion density, $n_{\rm de}$, of 1.5$\times 10^5$~cm$^{-3}$ as
indicated by the two vertical arrows. The N$_2$H$^+$ (dashed line) and
NH$_3$ (dotted line) abundances are found to increase by one to two
orders of magnitude over the region where CO is depleted, whereas HCN
(dashed-dotted line) is seen to follow the CO abundance, dropping by a
factor 5-10 over the same region.}\label{chemmodel}
\end{figure}

These trends are illustrated with a chemical model for the L483
envelope using the chemical network and approach described in
\cite{doty02,doty04} and adopting the physical structure from
\cite{jorgensen02} with CO depleted in the zone constrained by the
single-dish and interferometer observations above. As shown in
Fig.~\ref{chemmodel} this gives exactly an increase in N$_2$H$^+$ and
NH$_3$ over the region where the CO is depleted. Note that although
the CO abundance goes up again in the outermost part of the envelope
and the NH$_3$ and N$_2$H$^+$ abundances consequently drop, this
occurs at distances of $\approx$~8000~AU corresponding to 40\arcsec\
from the compact source in Fig.~\ref{momentone}, which is close to the
boundary of the N$_2$H$^+$ emission. This is, however, probably not the
main reason for the outer edge to the N$_2$H$^+$ emission: it is more
likely caused by the critical density of the N$_2$H$^+$ 1--0
transition which is $\approx 10^5$~cm$^{-3}$, a density which is also
reached at this distance from the central source. The N$_2$H$^+$
emission is therefore not sensitive to the outer region with low
density material where CO is undepleted.

Of course a detailed radiative transfer model for the N$_2$H$^+$ and
NH$_3$ emission could address some of these issues in more detail,
such as it was done for the CO emission. The problem is, however, that
the spherical model clearly breaks down for the scales probed by the
N$_2$H$^+$ emission. In fact the asymmetry may well be the reason that
N$_2$H$^+$ is seen in so clearly distinct zones away from the central
protostar in the L483 case, since N$_2$H$^+$ becomes marginally
optically thick toward the source center in an entirely spherical
core. Therefore N$_2$H$^+$ will not be sensitive to the innermost region
where the CO comes off the dust grains in the symmetric case.

\subsection{UV irradiation of outflow cavity walls}\label{cnchem}
\begin{figure}
\resizebox{\hsize}{!}{\includegraphics{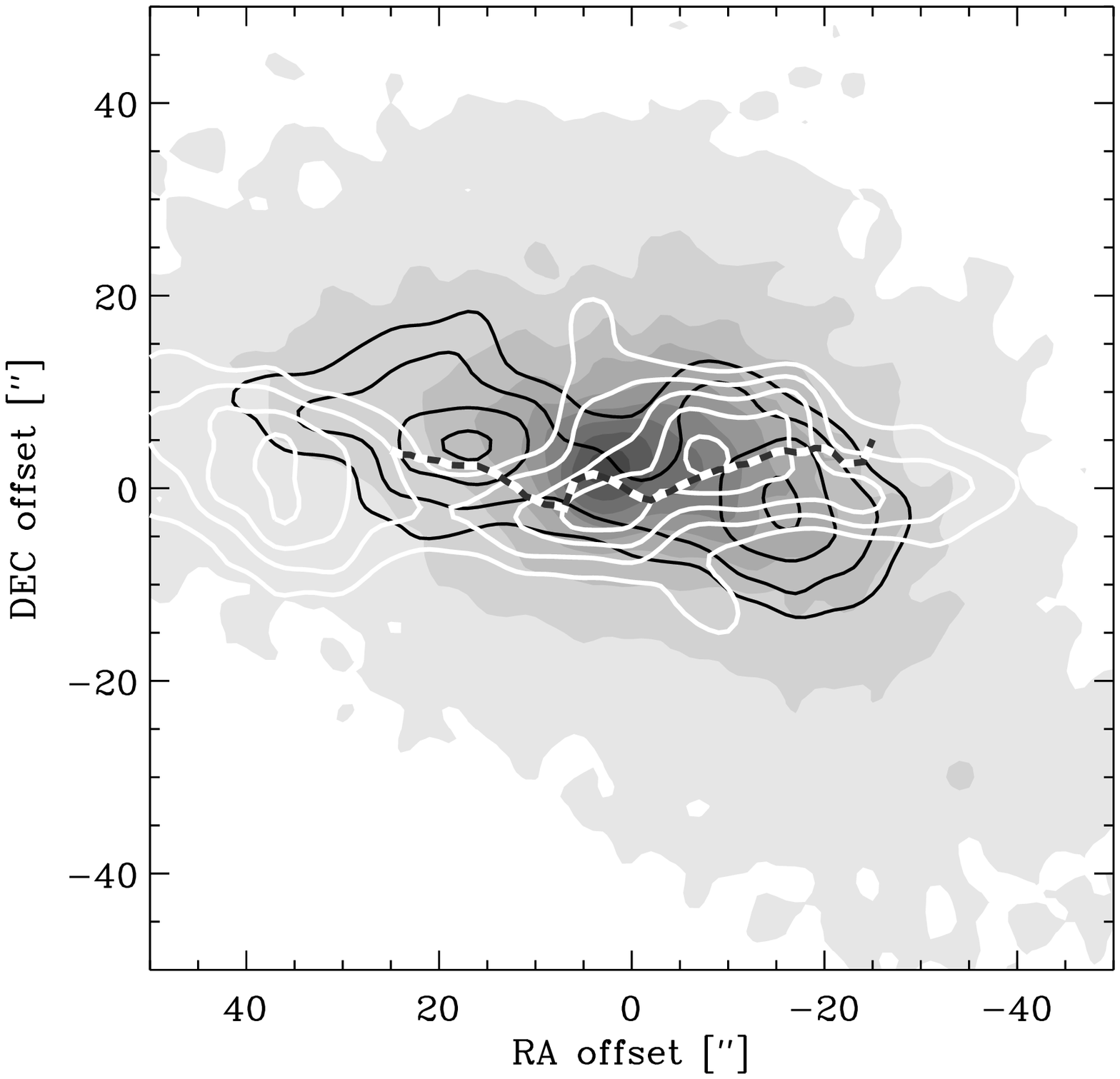}}
\resizebox{\hsize}{!}{\includegraphics{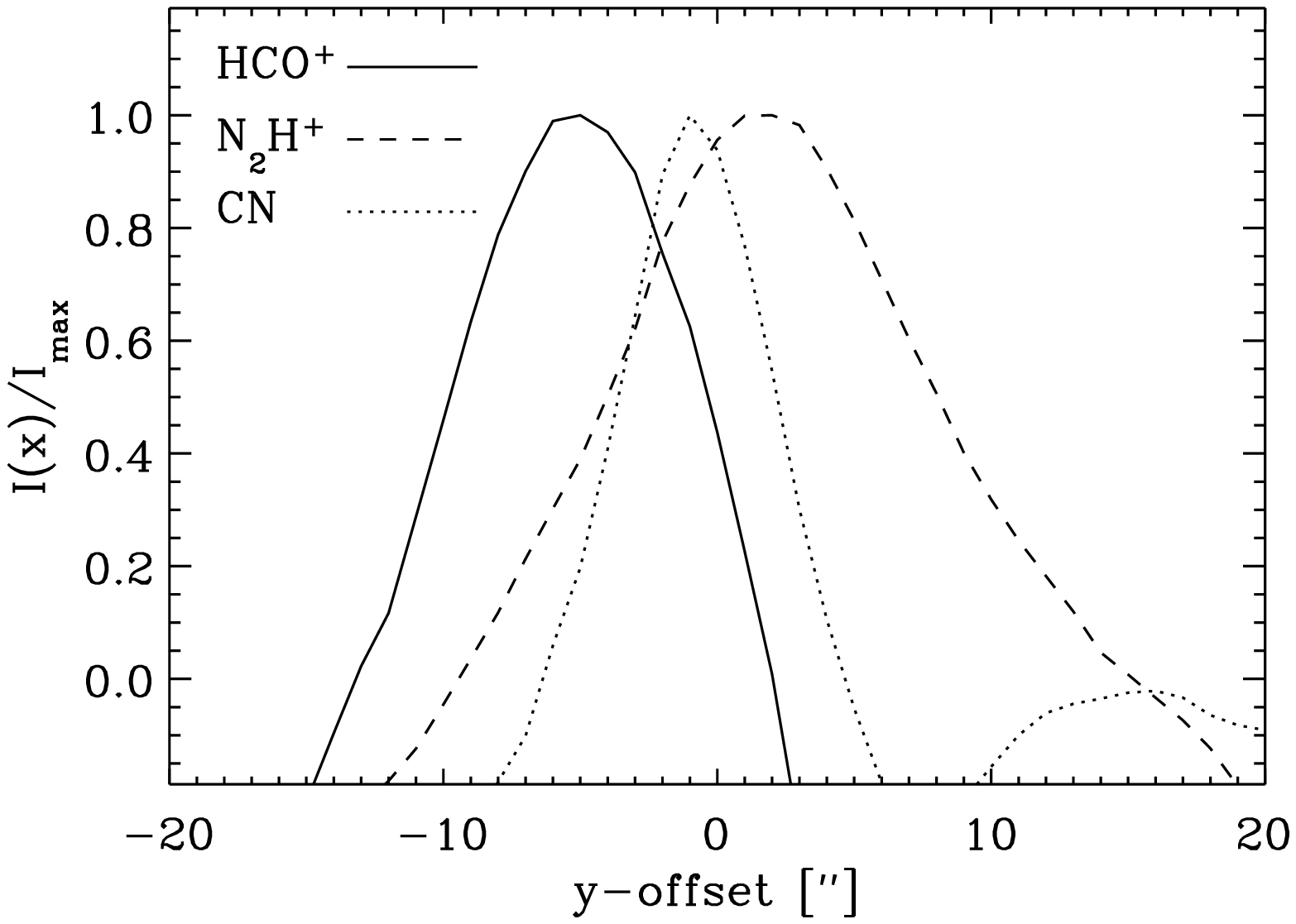}}
\caption{Comparison between HCO$^+$, N$_2$H$^+$ and CN
emission. \textit{Upper panel:} SCUBA 450$\mu$m image with
HCO$^+$ emission and N$_2$H$^+$ emission as white and black solid line
contours, respectively, and the vertical centroid of the CN emission
overplotted (black/white dashed line). \textit{Lower panel:} emission
of the three species in the vertical direction averaged over the
eastern outflow lobe (offsets between 10 and 25\arcsec\ in the
coordinate system aligned with the flattened direction of the core
(dashed lines in Fig.~\ref{nthp_co}).}\label{cn_hcop_n2hp}
\end{figure}

HCO$^+$ is found to be most prominent towards the outflow as is
typically seen at large and small scales
\citep[e.g.,][]{hogerheijde98}, whereas the N$_2$H$^+$ emission probes the
cold part of the quiescent cloud. As illustrated in
Fig.~\ref{cn_hcop_n2hp}, the CN emission appears in a boundary region
between HCO$^+$ and N$_2$H$^+$, i.e., between the outflowing and quiescent
cloud material. A possible explanation for this ``borderline
property'' of the CN emission could be that CN is enhanced in the
walls of an outflow cavity probed by the HCO$^+$ emission.

\cite{paperii} found a clear trend between the CN abundances and CS/SO
abundance ratio and suggested that this might reflect the strength of
ultraviolet radiation in the protostellar envelope. L483 is the source
in the sample of \cite{paperii} with the highest ratio of CN
abundances in the inner region relative to the outer region of the
envelope (i.e., [CN]$_{32}$/[CN]$_{10}$, where [CN]$_{32}$ and
[CN]$_{10}$ are the CN abundances inferred from the high excitation
3--2 lines at 340.24~GHz and low excitation 1--0 lines at 113.49~GHz,
respectively). For most sources in the sample this ratio is less than
1 indicating that the CN abundances are enhanced predominantly in the
low density material in the outermost region of the envelope. For
L483, the ratio is 2.5 indicating a higher CN abundance in the denser
material. If the CN abundances are probing the strength of the UV
field, this argues in favor of an internal source of the UV radiation
in the case of L483.

Such correlations between the CN abundances and UV field have been
suggested to be present in gaseous disks around more evolved class
I/II objects \citep{dutrey97,qi01,bergin03,vanzadelhoff03,thi04} which
are known to be strong emitters of UV radiation resulting from the
ongoing accretion. In the deeply embedded stages, however, the UV
radiation cannot escape far out into the envelope due to the large
extinctions. This will prevent direct detection of any UV flux from
the central star-disk system, but this emission may still be dominant
at small scales close to the central protostar. It is therefore of
high interest to search for specific chemical probes that may be used
to address the impact of UV in the inner envelope and thereby
potentially also the ongoing accretion.

The UV radiation field may furthermore be of large importance in
cavities cleared by outflows. \cite{spaans95}, for example, suggested
that the presence of high $J=6-5$ CO lines toward embedded protostars
is related to heating of material in the outflow cavity walls by the
UV radiation escaping from the central protostar. The near-infrared
nebulosity toward L483 (Fig.~\ref{tomass} and \cite{fuller95})
suggests that such a cavity is present. Alternatively the enhanced UV
radiation could be a result of the shocks associated with the L483
outflow which are directly probed by the H$_2$ emission
\citep{hodapp94,buckle99}. \cite{molinari01} suggested that
recombination in the post-shocked gas could illuminate outflow
cavities behind Herbig-Haro objects producing a PDR resulting in
diffuse C$^+$ emission seen by ISO \citep{molinari01,molinari02}.

To get a handle on this possibility a simple estimate of the 
extent of the CN emitting region can be made: the CN emission is
unresolved in its transverse direction, meaning that the extent of the
emission must be $\lesssim$ 5\arcsec. In the model for the L483
envelope from \cite{jorgensen02} the density at 20\arcsec\ from the
protostar is $n({\rm H_2})=3\times 10^{5} {\rm cm}^{-3}$. Introducing
the extent of the CN emission as $l \lesssim 5\arcsec$, an upper limit
to the column density of the CN emitting region can be estimated:
\begin{equation}
N_{\rm H_2} \lesssim 4\times 10^{21} {\rm cm}^{-2}\left(\frac{n({\rm
H_2})}{3\times 10^5 {\rm
cm}^{-3}}\right)\left(\frac{l}{5\arcsec}\right)
\end{equation}
or a visual extinction, $A_{\rm V} \lesssim 4$. Models of photon
dominated regions \citep[e.g.,][]{jansen95,sternberg95} indicate that
this is the extinction range over which CN would be enhanced
predominantly due to photodissociation of HCN. Alternative
explanations such as excitation effects appear less likely, since CN
is the only molecule showing this peculiar morphology.

\begin{figure}
\resizebox{\hsize}{!}{\includegraphics{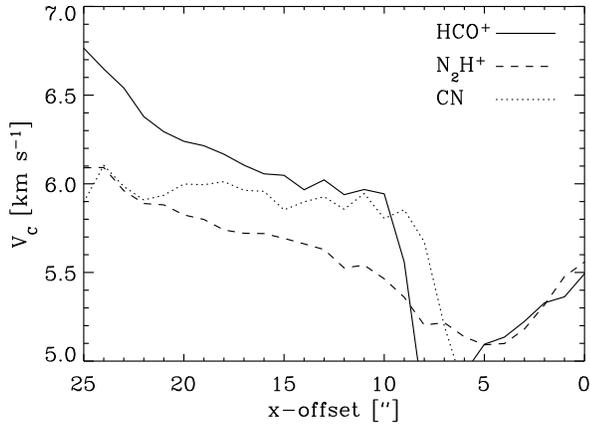}}
\caption{Velocity field in HCO$^+$, N$_2$H$^+$ and CN emission:
average centroid velocity of the three species in strips perpendicular
to the outflow axis across the eastern part of the
core.}\label{cn_hcop_n2hp_vel}
\end{figure}

If CN in fact is tracing the interface region between the outflow and
the quiescent core probed by HCO$^+$ and N$_2$H$^+$, respectively, this
should also be reflected in the observed velocity field of the three
species. As discussed above the velocity field probed by N$_2$H$^+$ shows
a red-blue spatial asymmetry aligned with that seen in HCO$^+$,
suggesting that the ``quiescent'' material probed by N$_2$H$^+$ is
being accelerated by the outflow. Fig.~\ref{cn_hcop_n2hp_vel} compares
the velocity field in the eastern lobe of the three species along the
flattened direction of the core (dashed line in
Fig.~\ref{nthp_co}). The HCO$^+$ does indeed show the largest velocity
gradient across the region with less steep increases in CN and N$_2$H$^+$
toward the lower densities in the core. Again CN is seen to be
intermediate between the outflowing and quiescent material in support
of the suggested scenario.

\section{Conclusions}
We have presented an analysis of the physical and chemical properties
of the class 0 protostar \object{L483-mm} on 500-1000~AU scales. This
paper demonstrates how the 1D envelope model derived from single-dish
continuum and line data can be applied to the interpretation of the
millimeter interferometer observations. The main conclusions are as
follows:

\begin{enumerate}
\item The continuum emission at 3~mm is well-fit by the envelope model
from \cite{jorgensen02}. In contrast to the case for other recently
studied class 0 objects (e.g., \object{NGC~1333-IRAS2A},
\object{L1448-C} and \object{IRAS~16293-2422};
\cite{n1333i2art,hotcorepaper}), an additional point source (e.g., a
disk) is not needed to fit the observed continuum emission. Assuming
an optically thin 30~K disk, this puts an upper limit to the disk mass
of 0.04~$M_\odot$.
\item The C$^{18}$O emission is found to be well-described in a ``drop
abundance'' model \citep[][]{coevollet,paperii} where the abundance is
high in the outermost regions (densities lower than
1.5$\times 10^5$~cm$^{-3}$) and innermost (temperatures higher than 40~K)
parts of the envelope but frozen out in between.
\item N$_2$H$^+$ and C$^{18}$O are found to be clearly anti-correlated, with
C$^{18}$O centered on the central continuum source whereas N$_2$H$^+$ has two
distinct peaks away from the central star. This is interpreted as the
combined effects of CO freeze-out at low temperatures and destruction
of N$_2$H$^+$ by reactions with CO. The N$_2$H$^+$ emission resembles
that of NH$_3$, previously reported by \cite{fuller00}, which can be
understood if NH$_3$ is destroyed through reactions with C$^+$ when
the CO abundance is high, and as clearly illustrated by a chemical
model for the L483 envelope.
\item HCN and CS emission probe the dense material close to the
central protostar. A velocity gradient perpendicular to the outflow
propagation direction, also recognized in previous C$_3$H$_2$
observations by \cite{park00}, is interpreted as rotation around
1~$M_\odot$ central protostar.
\item CN is found to trace a boundary between the quiescent material
probed by N$_2$H$^+$ and the larger scale outflow seen in HCO$^+$. A
possible explanation is that CN probes material in the outflow cavity
walls (seen as an infrared nebula in 2MASS $K_{\rm s}$ images) where
its abundance is enhanced as a result of UV irradiation. This is
further supported by the observed velocity field: the gradient
introduced by the outflow is seen in both HCO$^+$, CN and N$_2$H$^+$, with
HCO$^+$ showing the largest velocities relative to the systemic
velocity followed by CN (in the cavity walls) and N$_2$H$^+$ (in the core
material). This also suggests that a clear interaction between the
outflowing and quiescent material around L483 is taking
place. Possibly the outflow is in the process of dispersing the
protostellar core, causing its characteristic asymmetric shape.
\end{enumerate}

This paper illustrates the potential of high-resolution millimeter
interferometer observations for addressing the spatial variations in
the chemistry around protostellar objects. In particular, future high
resolution observations from the SMA, CARMA and eventually ALMA will
confirm or reject the drop abundance model through imaging of the
radial structure of the emission from high excitation lines of, e.g.,
C$^{18}$O, probing more uniquely the higher temperatures and densities
in the envelope. Also high resolution observations of high excitation
lines of CN will make it possible to further address the importance of
UV radiation in the inner envelope close to the central protostar and
confirm the relation between CN abundances and strength of the UV
field. In this context direct imaging of lines of, e.g., atomic carbon
in the high frequency windows will also be important to understand the
specific chemical effects. Such observations will become feasible with
the SMA and ALMA. An additional test would be deep mid-infrared
Spitzer observations: for example, deep imaging of the thermal emission
in the outflow cavities and searches for established probes of the UV
field such as the presence and properties of specific emission lines
and PAH features, provide complementary information which could be
used to establish a more detailed 2D model of complex protostellar
sources such as L483.

\begin{acknowledgement}
Ewine van Dishoeck, Michiel Hogerheijde, Fredrik Sch\"{o}ier and Geoff
Blake are thanked for fruitful discussions and comments on the
manuscript. Michiel Hogerheijde and Floris van der Tak are further
acknowledged for making their radiative transfer code publically
available. Steve Doty is thanked for the use of his chemical
code. This research was funded by a NOVA network 2 Ph.D. stipend.
\end{acknowledgement}

\bibliographystyle{aa}
\end{document}